\documentclass[11pt,letterpaper]{article}
\usepackage{jheppub}

\usepackage{graphicx}
\usepackage{bbm}
\usepackage{amsmath}
\usepackage{amssymb}
\usepackage{mathtools}
\usepackage{amsthm}
\usepackage{mathrsfs}
\usepackage{marvosym}
\usepackage{dsfont}
\usepackage[labelformat=simple]{subcaption}
\usepackage{xcolor}
\usepackage{braket}
\usepackage{cleveref}
\usepackage{comment}
\usepackage{float}
\usepackage{fancybox}
\usepackage{soul} 
\usepackage[skins,theorems]{tcolorbox}
\tcbset{highlight math style={enhanced,
		colframe=black,colback=white,arc=0pt,boxrule=1pt}}

\allowdisplaybreaks

\definecolor{dark-gray}{gray}{0.20}
\definecolor{gray}{gray}{0.30}
\definecolor{light-gray}{gray}{0.80}
\definecolor{dark-red}{rgb}{0.7,0,0}
\definecolor{dark-green}{rgb}{0.1,0.4,0}
\definecolor{dark-blue}{rgb}{0.3,0.3,0.7}
\definecolor{light-blue}{rgb}{0.8,0.8,1}
\definecolor{swamp}{RGB}{240, 199, 197}

\def\be{\begin{equation}}
	\def\ee{\end{equation}}
\def\bea{\begin{eqnarray}}
	\def\eea{\end{eqnarray}}

\newcommand{\SL}{\mathrm{SL}}

\hypersetup{
	colorlinks=true,
	linkcolor=dark-blue,
	citecolor=dark-red,
	urlcolor=dark-green,
	linktoc=page
}

\theoremstyle{remark}

\newtheoremstyle{named}{}{}{\itshape}{}{\bfseries}{.}{.5em}{#3}
\theoremstyle{named}

\title{\centering  End-of-the-World Singularities: \\ The Good, the Bad, and the Heated-up}

\author{Jos\'e Calder\'on-Infante,$^1$ Gongrui Cheng,$^2$  Alvaro Herr\'aez,$^3$ Thomas Van Riet$^2$}

\affiliation{$^1$Walter Burke Institute for Theoretical Physics, \\ California Institute of Technology, Pasadena, CA 91125, USA}
\affiliation{$^2$ Institute for Theoretical Physics, K.U.Leuven, Celestijnenlaan 200D, B-3001 Leuven, Belgium }
\affiliation{$^3$Max-Planck-Institut f\"ur Physik, Boltzmannstrasse 8, 85748 Garching bei M\"unchen, Germany}

\emailAdd{joseci@caltech.edu, gongrui.cheng@kuleuven.be, aherraez@mpp.mpg.de, thomas.vanriet@kuleuven.be}

\abstract{We revisit codimension-one End-of-the-World curvature singularities that drive scalars to infinite distance in field-space and have appeared in the context of dynamical cobordisms. We confront them with Gubser's horizon and potential criteria and with the Maldacena--Nu\~nez criterion. Moduli-space flows do not admit a near-extremal horizon generalization. Still, they satisfy Gubser's potential criterion and, in representative string realizations, the Maldacena--Nu\~nez criterion in ten dimensions. Together with an explicit uplift of this type of solution to a consistent string theory background, this suggests that such singularities should not be discarded. For flows with non-trivial scalar potential, we argue that the fate of the singularity is tied to the infinite-distance limit probed near the singularity. The Klebanov--Tseytlin and Klebanov--Strassler solutions illustrate that a modification that obstructs or modifies the field excursion should not be understood as a UV-resolution of the original singularity. We show that EFT strings and D7-branes fail Gubser's potential criterion despite having a sensible UV completion. Motivated by this, and inspired by dynamical cobordisms, we propose a novel criterion that bounds the divergence of the Ricci scalar as the flow explores infinite distance in field-space. Our criterion can be viewed as a geometrization Gubser's one that, while capturing all examples accepted by the latter, also admits EFT strings and D7-branes. Both criteria reject the massive Type IIA strong coupling End-of-the-World singularity. Finally, we analyze black D$p$-branes reduced to codimension one as representatives of flows that admit near-extremal generalizations, and find an exponential relation between temperature and field-space distance. This suggests a finite-temperature extension of the Distance Conjecture for dynamical cobordisms.
}

\begin{document}
	
\makeatletter
    \let\old@fpheader\@fpheader
	\renewcommand{\@fpheader}{ \vspace*{-0.1cm} \hfill CALT-TH 2026-013 \\ \vspace*{-0.1cm} \hfill MPP-2026-28}
	\makeatother
	
	\maketitle
	\setcounter{page}{1}
	\pagenumbering{roman} 
 
\pagenumbering{arabic} 
\section{Introduction}

The Cobordism Conjecture of \cite{McNamara:2019rup} states that the space of configurations in a quantum gravity theory belongs to the trivial cobordism class. A consequence of this is that, at least at the topological level, any such configuration should admit the introduction of a boundary ending spacetime. String theory offers concrete arenas to test this statement. Several works \cite{Buratti:2021yia,Buratti:2021fiv,Angius:2022aeq,Blumenhagen:2022mqw,Angius:2022mgh,Blumenhagen:2023abk,Calderon-Infante:2023ler,Angius:2023xtu,Huertas:2023syg,Angius:2023uqk,Angius:2024zjv,Huertas:2024mvy,Angius:2024pqk} suggest that these cobordisms to nothing are oftentimes dynamically realized as on-shell supergravity solutions with a spacetime singularity. These solutions have co-dimension one in the effective field theory description (typically after compactification) and may include a localized \emph{cobordism defect} at the singularity. They are often referred to as \emph{End-of-the-World branes} (ETW-branes).\footnote{See also \cite{Apers:2025pon,Apers:2026lgi, Raucci:2026fzp} for recent discussions of ETW-branes in different contexts and \cite{Nevoa:2025xiq} for recent work examining the elementary generators of Quantum Gravity cobordism classes and their tensions.} These dynamical cobordisms come with an appealing feature: as the solution approaches a spacetime singularity at finite spacetime distance, the scalars driving the flow traverse infinite distance in field space \cite{Buratti:2021fiv,Angius:2022aeq}. 

However, caution is required because an EFT singularity may simply signal an unphysical background (for example, a white-hole singularity). It is thus unclear whether any dynamically generated co-dimension one singularity that explores infinite distance in field space should be read as a physical cobordism defect. Our aim is to shed some light on this point by confronting these solutions with known regularity criteria that, while conjectural, have been extensively tested: Gubser’s \cite{Gubser:2000nd} and Maldacena–Nu\~{n}ez (MN) criteria \cite{Maldacena:2000mw}.

Gubser’s criterion \cite{Gubser:2000nd} requires the scalar potential to be bounded above as the scalar flows to the singularity, and  was formulated as a \emph{necessary} condition for a \emph{good} singularity. The key motivation was that acceptable singularities should admit a near-extremal deformation with a horizon that cloaks the singularity, recovering the original singular solution in the extremal limit. Singularities that posses such near-extremal generalizations were proven to have a potential that is indeed bounded above along the solution  \cite{Gubser:2000nd}. Requiring singular solutions to have such near-extremal horizon generalization is, however, known to be too restrictive as a necessary condition. For instance, singular supergravity solutions dual to Coulomb branch vacua of $\mathcal N=4$ SYM violate it, while still presenting a scalar potential that is bounded above near the singularity. Additionally, the study of moduli-space flows (some admitting a UV-completion in string theory \cite{Calderon-Infante:2023ler} but never having near-extremal deformations with smooth horizons) and other well-known codimension-one flows with non-vanishing potential (such as Klebanov-Tseytlin \cite{Klebanov:2000nc} and Klebanov-Strassler \cite{Klebanov:2000hb}) that we analyze in this paper indicates that the existence of a near-extremal horizon solution can only serve as a sufficient criterion for a good singularity. One of the key findings that we present here is that also Gubser's restriction on the potential may be too strong as a \emph{necessary} one, since it is in tension with EFT string solutions \cite{Lanza:2020qmt,Lanza:2021udy, Lanza:2022zyg}, which are known to have well-behaved UV uplifts in string theory \cite{Marchesano:2022avb, Martucci:2022krl, Marchesano:2022axe, Marchesano:2023thx, Martucci:2024trp, Marchesano:2024tod, Hassfeld:2025uoy, Grieco:2025bjy, Monnee:2025ynn}. 

Another important criterion is the Maldacena-Nu\~{n}ez (MN) one \cite{Maldacena:2000mw}, which in its \emph{strong} form allows a singularity in the infrared region of a holographic geometry only if the metric component $|g_{00}|$ does not increase as the singularity is approached (in its \emph{weak} form it is enough that it is bounded above). The \emph{strong} form applies to singularities interpreted as IR endpoints of an AdS/CFT dual pair, whereas the \emph{weak} form was motivated to accommodate other cases that violate the strong form but are still believed to be well-behaved, such as D8 brane domain walls in Type IIA.

Even though these criteria and some generalizations (see, e.g. \cite{Charmousis:2010zz}) were mainly motivated by holography and are often framed for asymptotically AdS spaces, the local character of a singularity suggests that its admissibility should not depend on the far asymptotics (at least as long as they are regular). It is therefore natural to examine them not only in asymptotically AdS backgrounds but also in asymptotically flat ones. This is one of our main motivations, especially for scalar flows on moduli spaces of Minkowski vacua. 
Furthermore, this idea supports the formulation of a singularity criterion through a condition on the background locally around the singular locus (morally following MN \cite{Maldacena:2000mw}), instead of referring to the dynamics generating it (as is the case of Gubser's  criteria \cite{Gubser:2000nd}). In the context of dynamical cobordisms, the latter can be turned into a geometric condition on how the Ricci scalar diverges as a function of the field-space distance as the singularity is approached. This allowed us to formulate  geometric version of Gubser's necessary condition that fits all known examples, including the aforementioned EFT strings that seemed to be in tension with Gubser's  restriction on the potential.

Beyond elucidating further which singularities correspond to \emph{good} ETW-branes, it is natural to wonder whether finite-temperature deformations of the ETW-brane can encode features of the towers required by the Distance Conjecture \cite{Ooguri:2006in}. Note, however, that it is not clear that the tower that becomes light and is stable near an infinite distance limit in moduli space (as predicted by the Distance Conjecture \cite{Ooguri:2006in}) also becomes light and stable in the ETW flow that probes that same infinite distance point, due to interactions sourced by the dynamical field variations.  Still, it is tantalizing to consider whether the finite-temperature generalization may excite the would-be light modes to account for its thermodynamic behavior, in qualitative analogy with \emph{minimal} black holes \cite{Cribiori:2023ffn,Basile:2023blg,Basile:2024dqq,Herraez:2024kux,Herraez:2025clp,Aparici:2026xxx}. We will present heuristic arguments to motivate an extension of the distance conjecture for finite $T$:
\begin{equation}\label{Eequation}
E \sim e^{-\tilde{\alpha}\,\Delta_\phi}\,,    
\end{equation}
with $E\simeq m$ at zero temperature (for particles at rest), and thus $\tilde{\alpha}$ reducing to the usual order-one Distance Conjecture parameter; whereas $E\simeq T^c$ (where $c$ is an order-one constant) at finite temperature.

The rest of this paper is organized as follows. In section \ref{sec:EOM} we review the general formulae for codimension-one flows that explore infinite distances in fields space, as well as their finite temperature generalizations. The aforementioned singularity criteria and some results about dynamical cobordisms are also reviewed. In section \ref{sec:ModuliSpaceFlows} we apply this formalism to explore flows in moduli space (i.e. without a scalar potential). We present the general solution for asymptotically flat flow solutions, as well as particular examples of flows in AdS$_5\times S^5$ and AdS$_3\times S^3\times \mathbbm{T}^4$. We confront these with Gubser's and Maldacena-Nu\~{n}ez criteria, and present an explicit UV-completion to determine the regime of applicability of such criteria. We turn to cases with non-vanishing scalar potential in section \ref{s:Flowsoutsidemoduli}, where we discuss various examples of ETW singularities in light of the singularity criteria and present a novel one motivated by geometrizing Gubser's potential criterion in the context of dynamical cobordisms. In section \ref{s:FiniteT} we revisit the codimension-one solutions from black D$p$-branes reduced on the transverse sphere and their non-extremal generalizations, and analyze the exponential dependence of the temperature with the moduli space distance. We leave a summary of results together with the conclusions and outlook for section \ref{s:Conclusions}. Finally, Appendix \ref{sec:AppendixBlackDpbranes} includes details on the scaling of the temperature of black D$p$-branes (when reduced to codimension-one flows) with the scalar-field distance.

\section{Domain wall flows and Singularity Criteria}\label{sec:EOM}
In this section we present the codimension-one solutions that are subject to our analysis and review the different singularity criteria that we use to analyze them as well as some results about dynamical cobordisms.

\subsection{Ansatz and effective actions: General Formulae}\label{ss:AnsatzDomainWalls}
We are interested in co-dimension one brane-like solutions of the following Lagrangian density in $D$ dimensions:
\begin{equation}\label{action1}
\mathcal{L} = \mathcal{R} - G_{ij}(\phi)\partial_\mu \phi^i\partial^\mu \phi^j - 2 V(\phi)\,.
\end{equation}
Here $\mathcal R$ is the Ricci scalar, $G_{ij}(\phi)$ is the field-space metric, and $V(\phi)$ is the scalar potential.

Solutions ``without temperature" have the following Ansatz 
\begin{equation}
\label{eq:ansatzzerotemp}
ds^2_D = N(r)^2 dr^2 + g(r)^2 ds^2_k\,,
\end{equation}
where the slicing $ds^2_k$ is either Minkowski ($k=0$), de Sitter ($k=+1$) or anti-de Sitter ($k=-1$) with normalised curvature. We assume the scalars depend only on $r$. The corresponding equations of motion can be derived from the following Lagrangian:
\begin{align}
L= & \frac{g^{D-1}}{N}\left[(D-2)(D-1)\left(\frac{g'}{g}\right)^2 -G_{ij}\phi'^i\phi'^j\right]\nonumber \\ & + (D-1)(D-2) k N g^{D-3} -2g^{D-1}N V(\phi) \,,
\end{align}
where a prime denotes a derivative with respect to $r$.
The variable $N$ (lapse function) is a Lagrange multiplier enforcing the constraint
\begin{equation} \label{constraint1}
\left( \frac{g'}{N}\right)^2 = k+ \frac{g^2}{(D-1)(D-2)}\left(\frac{1}{N^2}G_{ij}\phi'^i\phi'^j -2 V \right)    \,.
\end{equation}
The remaining equations of motion are those for the scalars, which can be derived from varying the effective Lagrangian above with respect to $\phi^i$:
\begin{equation}
\frac{2}{g^{(D-1)}}\frac{d}{dr}\left(\frac{g^{D-1}}{N} G_{ij}\phi'^{j} \right) = \frac{1}{N}(\partial_iG_{kl})\phi'^k\phi'^l+2N\partial_iV \,,
\end{equation}
where $\partial_i$ denotes the derivative with respect to $\phi^i$.

For simplicity, in this work we will only consider solutions with Minkowski slicings ($k=0$). At finite temperature, such solutions are generalized by the following Ansatz:
\begin{equation}
ds^2_D = N(r)^2 dr^2 + g(r)^2 \left(-h(r)^2 dt^2 + d\vec{x}_{D-2}^2 \right)\,.   
\end{equation}
The variable $h(r)$ is sometimes called \emph{blackening factor}. To present the effective 1d action we define the new variable $A = g^{D-1}h$. The equations then descend from \cite{Kuperstein:2014zda}
\begin{equation}\label{eq:actionwithh}
L = \frac{A}{N}\left[\frac{D-2}{D-1}\left(\frac{A'^2}{A^2} -\frac{h'^2}{h^2}\right) -G_{ij}\phi'^i\phi'^j\right] -AN2V(\phi)\,. 
\end{equation}
The constraint equation now is
\begin{equation}\label{constraint2}
\frac{1}{N^2}\left[\frac{D-2}{D-1}\left(\frac{A'^2}{A^2} -\frac{h'^2}{h^2}\right) -G_{ij}\phi'^i\phi'^j\right] +  2V(\phi) =0\,. 
\end{equation}
Note that we can always integrate one equation
\begin{equation}\label{v-equation}
\frac{Ah'}{N h} = v\,,    
\end{equation}
with $v$ some constant which is related to the temperature $T$ as\footnote{We work in units for which Boltzmann's constant equals 1.} 
\begin{equation}
T=\frac{v}{2\pi}g^{2-D}|_{r=r_0}\,.    
\end{equation}
In the gauge $N=A$ we can explicitly integrate this to
\begin{equation}
h(r) = \exp(v r  + w)\,,
\end{equation}
with $w$ an integration constant which can be fixed to zero by rescaling time.

\subsection{Singularity Criteria: Gubser and Maldacena-Nuñez} \label{ss:singularitycriteria}
Here we review previous attempts to the formulation of a criterion that diagnoses which curvature singularities of codimension-one solutions of the form described in section \ref{ss:AnsatzDomainWalls} can be allowed in gravitational EFTs.

\subsubsection*{Gubser's Criterion}

Gubser’s criterion, as originally formulated in \cite{Gubser:2000nd}, requires the scalar potential of a codimension-one singular solution to be bounded above as the singularity is approached for the latter not to be \emph{bad}. This is, it was presented as a \emph{necessary} (but not sufficient) condition for a \emph{good} singularity, and we refer to it as \textbf{Gubser’s \emph{potential} criterion}.\footnote{Throughout the text,  we slightly abuse the notation and refer to singularities as \emph{good} according to Gubser's \emph{potential} criterion whenever they fulfill this necessary condition. This simply means that they cannot be ruled out by the criterion and thus can potentially be \emph{good}, but note this is not guaranteed only by passing Gubser’s \emph{potential} criterion.} One of the most appealing features of this criterion in our context is that its formulation relies solely on local quantities near the singularity. This is an avatar of the idea that the local character of a singularity suggests that its admissibility should not depend on the far asymptotics. Even though this criterion was certainly formulated and is more directly applicable to asymptotically AdS solutions, this property allows us to apply it also to asympotically Minkowski backgrounds, or even to solutions with more cumbersome asymptotics.

A central motivation behind the idea of forbidding positive divergent scalar potentials was that acceptable singularities should admit a near-extremal deformation whose horizon cloaks the singularity, recovering the original singular solution in the extremal limit. We refer to this as \textbf{Gubser’s \emph{horizon} criterion}. It  was mainly rooted in the idea that, in AdS/CFT, the dual CFT configurations could support finite (arbitrarily small) temperatures. In fact, it was proven in \cite{Gubser:2000nd} that for singularities with AdS asymptotics that posses such near-extremal generalizations the potential is always bounded above. This links both criteria, as satisfying the latter guarantees satisfying the former, but the converse is not guaranteed. Gubser’s \emph{horizon} criterion is, however, too restrictive. As already noted in \cite{Gubser:2000nd}, singular supergravity solutions dual to Coulomb branch vacua of $\mathcal N=4$ SYM violate it, since any infinitesimal temperature drives the theory to the origin of the moduli space. Yet, these vacua are acceptable in the CFT and their dual singularities are expected to be \emph{good}. Importantly, these Coulomb branch solutions satisfy the (necessary) \emph {potential} criterion. This is thus consistent with treating Gubser's \emph{potential} criterion as a \emph{necessary} condition, while Gubser's \emph{horizon} criterion must be seen at most as a \emph{sufficient} one.

It is natural to wonder whether the \emph{horizon} criterion could be relaxed to a condition that also admits singularities that can be cloaked behind a horizon with a minimum finite critical temperature, as opposed to the case described above, in which that critical temperature can be continuously taken to zero in the near-extremal solution to recover the singular solution. As we will review in section \ref{s:Flowsoutsidemoduli}, this does not seem to provide a reasonable diagnostics, and also turns out to make the decoupling between the local tests and the asymptotics much more complicated.

\subsubsection*{Maldacena-Nu\~{n}ez Criterion}
The Maldacena-Nu\~{n}ez (MN) criterion \cite{Maldacena:2000mw} was originally motivated as a way to rule out singularities that cannot be understood as the IR endpoint in AdS/CFT. It was therefore formulated as a \emph{necessary} condition for a \emph{good} singularity. In its \emph{strong} version, it states that singularities in which the ten-dimensional Einstein-frame metric component $|g_{00}|$ does not grow as the singularity is approached are \emph{good}. The \emph{weak} version allows for growing $|g_{00}|$ as long as it remains bounded above.

The heuristic motivation behind it is the idea that for an excitation near the singularity in the bulk to be interpreted as the dual of some CFT state in the IR, it is expected that at fixed proper energy such excitation has a lower and lower red-shifted energy at the boundary. This is guaranteed as long as $|g_{00}|$ does not increase as we approach the singularity (i.e. the IR), as required by the \emph{strong} version; and the energy of the perturbation at least remains finite in energy if $|g_{00}|$ is bounded, as required by the \emph{weak} version. As opposed to Gubser's criterion, this is a purely geometric diagnostic that does not require knowledge about the dynamics generating it. This is, it does not require knowledge about which part of the contribution to the on-shell action comes from the scalar potential, and thus depends only on the background and not the generating dynamics. On the other hand, this criterion applies directly to the ten-dimensional metric. Starting from an effective action in $D<10$, applying it requires certain knowledge about the UV-completion. Furthermore, the uplift to ten dimensions in string theory may depend on the chosen duality frame, as it changes the geometric interpretation of the scalar fields. 

\subsection{Dynamical Cobordisms and End-of-the-World Singularities} \label{ss:dynamical-cobordisms}

For later convenience, we would like to introduce some of the results and terminology that has appeared in the literature in the context of so-called dynamical cobordism backgrounds and the scaling relations satisfied near the ETW singularities within them. We will mainly focus on summarizing the results in \cite{Buratti:2021yia,Buratti:2021fiv,Angius:2022aeq}, while referring the reader to \cite{Blumenhagen:2022mqw,Angius:2022mgh,Blumenhagen:2023abk,Calderon-Infante:2023ler,Angius:2023xtu,Huertas:2023syg,Angius:2023uqk,Angius:2024zjv,Huertas:2024mvy,Angius:2024pqk} for other interesting works on this topic.

\medskip

The term \emph{dynamical cobordism} \cite{Buratti:2021yia} makes reference to codimension-one solutions like the ones discussed in section \ref{ss:AnsatzDomainWalls} and emphasizes the expectation that these backgrounds are the EFT dynamical realization of the cobordism processes predicted by the Cobordism Conjecture \cite{McNamara:2019rup}. As argued in \cite{Buratti:2021fiv}, those describing \emph{cobordisms to nothing} feature an ETW singularity as the ones that are subject to this work. By studying a number of string-theoretic examples, it was found in \cite{Buratti:2021fiv,Angius:2022aeq} that these ETW singularities explore an infinite-distance limit in field space and that the Ricci scalar satisfies the scaling relation
\begin{equation} \label{scaling-relation}
    |R| \sim e^{\delta\, \phi} \quad {\rm as } \quad \phi\to\infty \, ,
\end{equation}
where $\delta > 0$ is the so-called critical exponent and $\phi$ represents the canonically normalized field that probes infinite field-space distance along the flow. This behavior is satisfied by all known top-down examples of codimension one singularities exploring an infinite-distance limit, as it is intimately related to the ubiquitous appearance of exponential potentials in string theory in these regimes. Indeed, it was argued in \cite{Angius:2022aeq} that ETW singularities satisfying \eqref{scaling-relation} are generated by potentials that asymptotically behave as 
\begin{equation} \label{dynamical-cobordism-potential}
    V \sim - c\, a \, e^{\delta \, \phi} \, , \quad \delta=2\sqrt{\frac{d-1}{d-2} (1-a)}  \, ,
\end{equation}
where both $c$ and $a$ are constants satisfying $c>0$ and $a<1$, respectively. More precisely, $a(\phi)=V/(V-T) \leq 1$ (with $V$ the potential and $T$ the kinetic energy of $\phi$) was used in \cite{Angius:2022aeq} as a convenient way of parameterizing potentials near the singularity. It was then shown that, whenever $a(\phi)$ converges to a constant $a<1$ as $\phi \to \infty$, the scaling relation in \eqref{scaling-relation} is recovered\footnote{Strictly speaking, the exponential scaling relation is not recovered in the fine-tuned case $\delta=\sqrt{2d/(d-2)}$ due to an accidental cancellation of the leading contribution to the Ricci curvature \cite{Angius:2022aeq}. In that case, the divergence of the Ricci scalar is rather bounded above by \eqref{scaling-relation}. This particular case can be included in our derivation below without changing the conclusion.} with $c$ appearing as an integration constant.\footnote{For completeness, let us discuss what happens when $a(\phi)$ does not converge to a constant $a<1$. First, the case $a(\phi) \to -\infty$ can be seen to produce super-exponentially growing potentials and lead to super-exponential scaling relations. Similarly, $a(\phi) \to 1$ leads to asymptotically negative potentials with sub-exponential behavior and to slow-roll solutions with sub-exponential scaling relations.} 

When $a \neq 0$, the kinetic energy of the scalar, $T$, and the potential, $V$, are asymptotically of the same order. This is analogous to the so-called \emph{scaling solutions} that appear in the study of cosmological backgrounds with exponential potential (see e.g. \cite{Halliwell:1986ja, Hartong:2006rt, Calderon-Infante:2022nxb}). On the other hand, the $a=0$ case---or, more precisely, $a(\phi)\to 0$---leads to $T \gg |V|$ near the ETW singularity. This case is analogous to that of \emph{kination}\footnote{In the older literature kination was instead refered to as `kinetic dominated scaling solutions'.} in cosmological scenarios (see e.g. \cite{Apers:2022cyl}). It leads to $\delta=\delta_{\rm crit} \equiv 2\sqrt{(d-1)/(d-2)}$ in \eqref{scaling-relation} and in fact encompasses any potential such that $|V| \ll \exp(\delta_{\rm crit} \, \phi)$ as $\phi \to \infty$. In other words, any potential satisfying this condition will admit an ETW singularity with $\delta=\delta_{\rm crit}$ in \eqref{scaling-relation} as a (locally) consistent solution. 

\medskip

The results we have just presented can be summarized as follows: Consider asymptotically exponential potentials of the form $V \sim \pm \exp{\gamma \phi}$, which are those leading to the scaling relation \eqref{scaling-relation}. We find the following classification of possible ETW singularities, characterized by the critical exponent $\delta$, depending on the sign of $V$ and the value of $c$:\footnote{The relations below crucially rely on flat slicings. Once one has AdS or dS sliced domain walls there are new scaling solutions \cite{Chemissany:2011gr,Angius:2022aeq}. An example of such curved ETW singularity arises from reducing Witten's bubble of nothing \cite{Witten:1981gj} on the circle.}
\begin{itemize}
    \item $V>0$ and $\gamma < \delta_{\rm crit}$: $\delta = \delta_{\rm crit}$ (kination).
    \item $V>0$ and $\gamma \geq \delta_{\rm crit}$: $\delta = \gamma$ (scaling).
    \item $V<0$ and $\gamma \leq \delta_{\rm crit}$: $\delta = \delta_{\rm crit}$ (kination) or $\delta = \gamma$ (scaling).
    \item $V<0$ and $\gamma > \delta_{\rm crit}$: No compatible ETW singularity.
\end{itemize}
Notice the lack of ETW singularity in the last case. A way of understanding this better comes from switching from domain wall to cosmological backgrounds, which amounts to changing the sign of $V$ (see e.g. \cite{Skenderis:2006fb}). What we are finding is that a scalar field is not allowed to evolve dynamically towards $\phi \to \infty$ with time if the potential is blowing up to $+\infty$ too fast. In other words, for any initial condition such that $\phi'_0>0$, the scalar will be eventually forced to turn around and to move towards negative values for $\phi$. On the other hand, the case with $V<0$ and $\gamma < \delta_{\rm crit}$ features two types of ETW singularities. This case happens e.g. for D$p$-branes with $0<p<7$ and, indeed, the D$p$-branes reduced to codimension-one solutions correspond to the scaling solution \cite{Chemissany:2011gr, Angius:2022aeq}.  The kination solutions in this case with non-constant string dilaton lift to the asymptotic regions of what are called ``type I" deformations of the known $p$-brane solutions in \cite{Lu:1996er}.

Before moving on, let us stress that all the discussion in this section is local around the ETW singularity. For instance, the classification above only refers to locally consistent solution to the equations of motion. Whether this ETW singularity can be embedded into a global solution is a non-trivial question that depends on the form of the potential beyond the $\phi \to \infty$ limit.

\section{ETW Singularities in Moduli Space Flows}
\label{sec:ModuliSpaceFlows}

In this section, we consider ETW-brane solutions in the moduli space of both asympotically Minkowski and AdS vacua with constant $V(\phi)=\Lambda$. First, we review domain wall flow solutions in moduli space in Section \ref{ss:moduli-space-flows}, showing that they indeed contain a codimension one singularity. To ellucidate whether this singularity is good or bad, we apply Gubser and Maldacena-Nu\~{n}ez criteria in Section \ref{ss:moduli-space-vs-criteria}. In Section \ref{ss:moduli-space-UV-complete}, we present examples of UV-complete ETW singularities in Minkowski moduli spaces of string theory.

\subsection{Domain wall flows in moduli space} \label{ss:moduli-space-flows}

The distance conjecture has been originally formulated for true moduli spaces, which are only expected to exist in presence of enough supersymmetry. Extensions of the distance conjecture to scalar fields that are not moduli, and hence are lifted by a potential, are more speculative and less understood \cite{Basile:2023rvm, Basile:2022zee, Debusschere:2024rmi,Mohseni:2024njl, Demulder:2024glx}. With this motivation in mind, we first investigate flows inside moduli space. This means that the potential $V(\phi)=\Lambda$ is constant. 

The equations of motion for moduli imply that they follow geodesic curves. This can be most easily understood by using the effective Lagrangian \eqref{eq:actionwithh} in the gauge $A=N$ where the scalar part of the Lagrangian becomes
\begin{equation}
L(\phi,\phi') = G_{ij}\phi'^i\phi'^j \,.   
\end{equation}
Hence, in the gauge $A=N$, $r$ defines an affine parametrization and we have 
\begin{equation} \label{eq:C}
 A=N\quad \longrightarrow\quad G_{ij}\phi'^i\phi'^j =C\geq 0 \,.
\end{equation}
In the general coordinate frame the scalars obey
\begin{equation}
G_{ij}\phi'^i\phi'^j=C\frac{N^2}{g^{2(D-1)}}>0 \, ,
\end{equation}
and \eqref{constraint1} simplifies to
\begin{equation}
    \left(\frac{g'}{N}\right)^2 =  k + \frac{g^2}{\ell^2} + \frac{C}{(D-1)(D-2)}\frac{1}{g^{2(D-2)}}\,,
\end{equation}
where we introduced the AdS length $\ell$ via
\begin{equation}
    \Lambda = -\frac{(D-1)(D-2)}{\ell^2}\,.
\end{equation}
Note that one can always take a gauge in which $g=r$ and then the above equation is solved for $N$ trivially. So there is always an explicit metric. Different expressions for the metric in other gauges can be found in for instance \cite{Bergshoeff:2008be}. 

Instead of solving for the general metric, let us simply demonstrate why there is a singularity in the bulk at zero $T$. For that keep in mind that $C>0$. Then the above constraint equation, in the gauge $N=1$ can be seen as a particle subject to an effective potential of the form
\begin{equation}
V_{\text{eff}}(g) =   - \left( k + \frac{g^2}{\ell^2} + \frac{C}{(D-1)(D-2)}\frac{1}{g^{2(D-2)}}\right)\,,
\end{equation}
with the total energy being zero. This potential is unbounded from below near $g\rightarrow 0$ where the equation simplifies to 
\begin{equation}\label{g'_approx}
 g'^2 \approx \frac{C}{(D-1)(D-2)}\frac{1}{g^{2(D-2)}}\,,   
\end{equation}
and the solution of this equation is singular:
\begin{equation}
g^{D-1} \approx r\sqrt{\frac{D-1}{D-2}\,  C}\, 
\end{equation}
Note that this singularity, to leading order, does not depend on $k$ or $\ell$. This singularity is timelike and hence naked. 

One can also track down the behavior of the scalar $\phi$ parametrizing the geodesic in moduli space as the singularity is reached.\footnote{By this we mean $\phi'^2 = G_{ij}\phi'^i\phi'^j$.} Again in the $N=1$ gauge, in which the coordinate $r$ is the transverse distance, one finds
\begin{equation} \label{eq:phi-ETW}
    \phi(r) = \mp \sqrt{\frac{D-2}{D-1}} \log r \,,
\end{equation}
  as we approach the singularity, which we have placed at $r=0$. We observe that the canonically normalized moduli diverges as the singularity is approached. In fact, we recover the ETW-brane scaling relations reviewed in section \ref{ss:dynamical-cobordisms} in the case in which the critical exponent $\delta$ vanishes.

These moduli space flows provide a nice dynamical realization of the infinite distance limits that are of interest for the Distance Conjecture. For instance, they were used in \cite{Calderon-Infante:2023ler} together with holographic bounds to provide a heuristic bottom-up argument for the conjecture in Minkowski moduli space. Even though this argument only relies on the behavior of the background far away from the singularity, having a singularity puts into question the validity of the background itself. In other words, this raises the question of whether this singularity is \emph{good} or \emph{bad}. In the next section, we use the currently existing criteria for \emph{good} vs. \emph{bad} singularities to address this question.

Before going on, let us remark something interesting about these solutions in relation to singularity criteria. Notice that the background looks exactly identical for any geodesic in moduli space. However, the microscopic origin of each modulus may be different. Thus, the UV-completion of the flow depends on the specific geodesic we choose. From this we learn that any UV-insensitive criterion such as Gubser's criteria or our criterion \eqref{eq:ETW-criterion} will yield the same answer for any flow in moduli space. In contrast, a UV-sensitive criterion such as the ones by Maldacena and Nu\~{n}ez may be able to tell different moduli space flows apart. Moduli spaces are thus an interesting setup to test the limitations of UV-insensitive criteria. For instance, we conclude that a UV-insensitive criterion that is \emph{complete} can only exist if all moduli space flows are good or bad. In Section \ref{ss:moduli-space-UV-complete}, we will present some examples of good moduli space flows, thus reaching the conclusion that the latter option is not possible. 

\subsection{Confronting moduli space flows with Gubser and Maldacena-Nu\~{n}ez} \label{ss:moduli-space-vs-criteria}

\subsubsection{Gubser's criteria}
\label{sss:Gubserformoduli}
First of all, let us quickly mention that any moduli space flow trivially fulfills Gubser's \emph{potential} criterion for any finite $V(\phi)=\Lambda$, since the potential is constant and always finite, even near the singularity. This means that these solutions cannot be ruled out on the grounds of Gubser's \emph{potential} criterion but, can we unequivocally conclude whether they are \emph{good} singularities?

To answer this question, let us explore the fate of moduli space flows under Gubser's \emph{horizon} criterion, which poses that a singularity is \emph{good} if it can be cloaked behind a near-extremal horizon. As we show next, these do not fulfill this criterion. Let us remark that this is a sufficient (but not necessary) condition for a \emph{good} singularity, so failing it does not automatically rule out the solution, as we subsequently discuss in section \ref{ss:moduli-space-UV-complete}.

 We will consider a single scalar, but the equations for a generic sigma model are a straightforward generalization. This analysis shows the horizon becomes singular and so we turn a timelike (naked) singularity  into a null singularity. The equations for the blackening factor and free scalar read
\begin{equation}
 \frac{h'}{h}=v\frac{N}{A}\,,\qquad \phi'=\sqrt{C}\frac{N}{A}\,,   
\end{equation}
with $C$ and $v$ two constants introduced in eqs. \eqref{eq:C} and \eqref{v-equation}, respectively. We find 
\begin{equation}
\phi = \frac{\sqrt{C}}{v} \ln{h} + \text{const.}\,,    
\end{equation}
which includes a finite integration constant. Hence, as $h$ approaches zero in order to develop a horizon, the scalar field diverges, indicating the singularity alluded to above. In the gauge $N=A$ the profiles of $\ln{h}$ and $\phi$ are linear. The solution for $A=g^{D-1}h$ can be obtained from the constraint equation
\begin{equation}
A'^2 = (v^2 +C\frac{D-1}{D-2}) A^2  -2\Lambda\frac{(D-1)}{(D-2)}\, A^4\,.    
\end{equation}
This can be solved explicitly, but for our purposes it suffices to notice that the solution interpolates between a curvature singularity for small $A$---which is null by the argument above---and AdS (or Minkowski) asymptotics for large $A$.  Hence, this solution transforms the original timelike singularity into a null singularity at finite temperature.\footnote{An alternative argument that shows this goes as follows: At the horizon the scalar derivative has to vanish in order to find a smooth solution for this case as explained in the appendix of \cite{Gubser:2000nd}, which implies $C=0$.}

Since moduli flows do not satisfy Gubser's \emph{horizon} criterion, we cannot use it to conclude that the singularities are \emph{good}. Nevertheless, this does not mean that they are \emph{bad}, it simply means they cannot be \emph{heated-up}. All in all, this UV-insensitive criterion is inconclusive in this case. 

\subsubsection{Maldacena-Nu\~{n}ez criteria}
\label{sss:MN}
In the following we turn to the Maldacena-Nu\~{n}ez criteria. These are  UV-sensitive in the sense that they require an uplift to ten dimensions. For this reason, we focus on two concrete moduli flow solutions with known uplifts: Gubser's flow in AdS$_5\times S^5$ and moduli flows inside AdS$_3\times S^3\times \mathbbm{T}^4$. We present a more general discussion in relation to the Emergent String Conjecture \cite{Lee:2019wij} at the end of \ref{ssss:GeneralMN}

\subsubsection*{Gubser's flow  in AdS$_5\times S^5$}
\label{sss:Gubserflow}
The simplest flow solution inside an AdS moduli space is known as the ``Gubser-Flow" \cite{Gubser:1999pk} and was constructed in 5d by simply turning on the dilaton in the $S^5$ compactification of IIB and truncating all other fields in the vacuum, which is a consistent truncation. The resulting system is described by \eqref{action1} with $G_{ij}(\phi)\partial\phi^i\partial\phi^j=(\partial\phi)^2$ and $V=\Lambda<0$. 
In the gauge $N=1$ we have
\begin{align}
\phi'^2 = \frac{C}{g^8}\,,\qquad
g'^2 = \frac{g^2}{\ell^2} +\frac{C}{12}\frac{1}{g^6}\,.
\end{align}
The boundary of AdS is then near $g\rightarrow \infty$ where the dilaton momentum goes to zero. A singularity will develop in the interior when $g$ becomes small $g\ll \ell$ and from \eqref{g'_approx} we find
\begin{equation}
 g(r) \approx \left(2\sqrt{\frac{C}{3}}\,\, r\right)^{1/4}\,,    
\end{equation}
Near the singularity at $r=0$ the dilaton field blows up like 
\begin{equation}
\phi \approx \mp\frac{\sqrt{3}}{2} \ln(r)
\end{equation}
Via S-duality we can restrict to the minus sign such that the string coupling vanishes near the singularity.  We note that (in Einstein frame) the singularity does pass the MN criterion since $|g_{00}|\sim r^{1/2}$. Since the $S^5$ is not flowing, the behavior of $g_{00}$ is the same in the 5d and the 10d solution. Concerning the fate of the singularity itself, the jury is still out on this one, but it is certainly not ruled out by any known criterion.

\subsubsection*{Moduli flows inside AdS$_3\times S^3\times \mathbbm{T}^4$}
\label{sss:AdS3flows}
Another simple AdS vacuum with a well-understood moduli space, dual to the exactly marginal operators, is AdS$_3\times S^3\times \mathbbm{T}^4$.  Following the notation and convenctions of \cite{Astesiano:2022qba} we can truncate the moduli space system down to the 3d lagrangian
\begin{equation}
\mathcal{L} = \mathcal{R}_3 -\tfrac{1}{2}(\partial\psi)^2- \tfrac{1}{2}(\partial\tilde{\phi})^2- \Lambda\,.
\end{equation}
Both $\tilde{\phi}$ and $\psi$ are moduli in a larger moduli space, but these two are particularly easy to understand as their singular flows are easily lifted to 10d.  The lift to 10d proceeds through the following equations: first we note that
\begin{equation}
\sqrt{2}\tilde{\phi} = \phi + \varphi\,,    \qquad e^{\phi-\varphi} = |\frac{Q_1}{Q_5}|\,,
\end{equation}
where $\phi$ is the 10d dilaton and $\varphi$ the torus volume modulus appearing in the 10d Einstein frame:
\begin{align}
    ds^2_{10} = &e^{\tfrac{1}{2}\varphi} \left(ds^2_{3} + ds^2_{S^3}\right)
    + e^{-\tfrac{1}{2}\varphi} \left(e^{\tfrac{1}{\sqrt2}\psi} \left(d\theta_1^2 +d\theta_2^2\right) + e^{-\tfrac{1}{\sqrt2}\psi} \left(d\theta_3^2 +d\theta_4^2\right) \right)\,.
\end{align}
Here we also see the modulus $\psi$ defined as a modulus that measures the relative size of the two 2-tori inside the 4-torus. The scalar combination orthogonal to $\tilde{\phi}$ defined through the difference  $\phi - \varphi$ is stabilised by the fluxes piercing the AdS$_3$ and $S^3$ factor. 

Before we analyze the flows, let us analyze the generic singularity in 3d, similarly to what we did in the previous subsection. We have $k=0, D=3$ and so in the gauge $N=1$ the IR singularity is governed by
\begin{equation}
 g(r)\approx (\sqrt{2C}r)^{1/2}\,.    
\end{equation}
To understand whether the singularity near $r=0$ is \emph{bad} we need to lift the solutions to 10d: 

\medskip

\begin{itemize}
\item{\textbf{$\psi$-flow:} 
Let us consider a solution in which only $\psi$ flows. Then, we have that near $r=0$, $ \psi(r)\approx -\ln(r)$ and so, near the singularity, the 10d metric reads:
\begin{align}
ds^2 \approx& \left(g_s\frac{|Q_5|}{|Q_1|}\right)^{1/2}\left(dr^2 + \sqrt{C}\, r \,\eta_{ij}\, dx^idx^j + d\Omega_3^2\right)\,+\nonumber\\
&  \left(g_s\frac{|Q_5|}{|Q_1|}\right)^{-1/2}\left(r^{-\frac{1}{\sqrt2}}\left(d\theta_1^2 +d\theta_2^2\right) + r^{\frac{1}{\sqrt2}} \left(d\theta_3^2 +d\theta_4^2\right)\right) \,.\nonumber
\end{align}
Therefore, the solution remains singular in 10d, but it survives the (strong) MN criterion. 
}
    \item{\textbf{$\tilde{\phi}$-flow:} We now consider a flow in which only $\phi$ runs. Near $r=0$ we have $\tilde{\phi} \approx -\ln(r)$, so 
\begin{equation}
    e^{2\phi} \approx |\frac{Q_1}{Q_5}|r^{-\sqrt2} \,,\qquad e^{2\varphi} \approx  |\frac{Q_5}{Q_1}|r^{-\sqrt2}\,.
\end{equation}
In the 10d string frame, the metric near the singularity reads
\begin{align}
ds^2\,  \approx\,  &r^{-\frac{\sqrt2}{4}}\left(\frac{|Q_5|}{|Q_1|}\right)^{1/4}\left(dr^2 +  \sqrt{C}\, r \,\eta_{ij}\, dx^idx^j + d\Omega_3^2 \right)+ \\&r^{\frac{\sqrt2}{4}}\left(\frac{|Q_1|}{|Q_5|}\right)^{1/4}\left(d\theta_1^2 +d\theta_2^2 + d\theta_3^2 +d\theta_4^2\right)\,.\nonumber
\end{align}
In the Einstein frame the $g_{00}$-component of the metric scales like
$|g_{00}| \sim  r^{-\frac{\sqrt2}{4}} r e^{-\phi/2}\sim r$,  and thus passes the (strong) MN criterion.} 
\end{itemize}
Thus, none of the flows can be ruled-out by the MN criteria.

\subsubsection*{Maldacena-Nu\~nez criterion and the Emergent String Conjecture}
\label{ssss:GeneralMN}

In $D$-dimensions, the time component of the metric in the moduli flows behaves as 
\begin{equation}
    |g_{00}^{(D)}| \sim g(r)^2 \sim r^{\frac{2}{D-1}} \quad \text{as } r\to 0 \, .
\end{equation}
For $D>2$, this power is always positive, so $|g_{00}^{(D)}|$ goes to zero as $r\to 0$. This behavior points in the right direction toward satisfying the MN criterion. However, we need to uplift to 10d to apply the latter. To make a general argument, we invoke the Emergent String Conjecture \cite{Lee:2019wij}, which states that, in some duality frame, the infinite-distance limit corresponds either to emergent weakly coupled strings or to a decompactification. Furthermore, the behavior of the string scale is irrelevant for the uplift to the 10d Einstein frame (as long as we stay within a perturbative regime). Therefore, we assume that the geodesic of the moduli flow leads to a decompactification to $(D+n)$-dimensions, where $(D+n)\leq10$.\footnote{If extended to M-theory, one can also include the $D+n=11$ case} The zero-zero component of the metric in 10d then behaves as
\begin{equation}
    |g_{00}^{(10)}| \sim e^{- 2 \upsilon \phi} |g_{00}^{(D)}| \, , \quad \upsilon = \sqrt{\frac{n}{(D-2)(D+n-2)}} \, .
\end{equation}
We are using the same conventions as in section \ref{ss:dynamical-cobordisms} (see \cite{Angius:2022aeq} for details), for which the scalar $\phi$ behaves as in \eqref{eq:phi-ETW} as we approach the singularity. We do not need the precise behaviour, we only need to know that the scalar is such that $\phi \to \infty$ when the $n$-dimensional volume blows up, and that $\upsilon >0$. This automatically implies that, when $r\rightarrow 0$ we find
\begin{equation}
    e^{- 2 \upsilon \phi} \to 0  \quad \, \Rightarrow \,\quad     |g_{00}^{(10)}| \to 0  \, ,
\end{equation}
such that the \emph{strong} MN criterion is satisfied.

\medskip

This result crucially relies on the assertion that all infinite distance limits lead to decompactification in some duality frame. In contrast, let us see what happens if we go to another duality frame in which instead of a decompactification we find that part of the internal space shrinks to zero size. In this case the $\phi$-dependent prefactor in front of $g_{00}^{(10)}$ will blow up, so the 10d uplift can fail to satisfy the MN criterion. To see when this can happen, recall that the modulus behaves as\footnote{The same expression with a minus sign corresponds to a decompactification.}
\begin{equation}
    \phi =\sqrt{\frac{D-2}{D-1}} \log r + \text{const.} \, ,
\end{equation}
such that,
\begin{equation}
    |g_{00}^{(10)}| \sim r^{2 \left(\frac{1}{D-1} - \sqrt{\frac{n}{(D-1)(D+n-2)}} \right)} \, .
\end{equation}
The MN criterion will then be fulfilled if the power in the equation above is non-negative. Interestingly, for $D>2$ this can only happen for $n=0$ (as in Gubser's flow) or $n=1$ (when a single extra dimension shrinks). Notice that the result drastically changes when comparing these two duality frames. This highlights the idea that the MN criterion seems to be frame-dependent, since one can consider different 10d frames that lead to the same EFT in $D$-dimensions. If it can be applied in the duality frame in which the singularity explores a decompactification limit, then all moduli space flows would pass the (strong) MN criterion as we saw above. 

\subsection{A UV-complete ETW singularity in moduli space and its generalization} \label{ss:moduli-space-UV-complete}

In the previous section, we saw how some conjectural singularity criteria are either inconclusive or suggest that ETW singularities in moduli flows are good. In what follows, we recall an example of such solution that can be UV-completed in string theory and present generalizations thereof. Hence, even though we cannot conclude that \emph{all} such ETW singularities are good, we establish that at least \emph{some} of them are.

The simplest example of a UV-complete ETW singularity in moduli space was introduced in \cite{Calderon-Infante:2023ler} (see Appendix A therein). The setup is Einstein gravity reduced on a circle with the Kaluza-Klein (KK) vector truncated, which leads to gravity coupled to a modulus (the radion). When the singularity explores the infinite distance limit in which the circle pinches off, the flow introduced above interestingly uplifts to $\mathbb R^{1,D-2} \times \mathbb C/\mathbb Z_k$, which only has a (co-dimension two) orbifold singularity that is resolved by the introduction of a twisted sector in string theory.\footnote{In general, one obtains a conical singularity. When a certain quantization condition is met, the deficit angle is that of an orbifold singularity. We refer the reader to \cite{Calderon-Infante:2023ler} for more details. This quantization of an integration constant in the UV completion is analogous to the case of D$p$-brane supergravity solutions, which admit a healthy UV-completion if the charge is properly quantized.} In this section, we would like to point out an interesting connection between the resolution of this singularity in string theory and the tower of light states of the Distance Conjecture in the limit explored by the ETW singularity. Furthermore, below we extend the computation to an $n$-torus and check what the solution uplifts to.

\medskip

In the adiabatic setup, i.e., when the moduli vary slowly, the relevant tower of states for the Distance Conjecture when a circle pinches off is given by winding modes of the string. When this infinite distance limit is realized dynamically as we approach an ETW singularity, the radion field is subject to large gradients and these winding states are highly unstable. From the $(D+1)$-dimensional perspective, this is apparent from the fact that these winding modes (i.e., macroscopic closed strings winding around the orbifold singularity) are allowed to shrink. From the $D$-dimensional perspective, where we forget about the angular coordinate around the tip of the cone, this means that these modes are very quickly attracted to the ETW singularity. For this reason, they cannot be interpreted as stable objects living in the dynamical cobordism geometry. This has been one of the main obstacles in connecting these dynamical realizations of infinite distance limit with the towers of states of the Distance Conjecture. However, the UV completion of this ETW singularity reveals a relation between this tower of (highly unstable) states and the resolution of the singularity in string theory. 

Recall that a string with winding number $\omega$ around a $\mathbb C/\mathbb Z_k$ orbifold singularity carries a $\omega$ mod $k$ conserved charge.\footnote{This is a $\mathbb Z_k$ charge because a string winding a multiple of $k$ times around the orbifold singularity is equivalent to a string winding around the origin in the embedding space $\mathbb C$ and is thus allowed to unwind.} As discussed above, a macroscopic string winding around the orbifold singularity is unstable against shrinking. For energy to be conserved, this will happen while closed string modes are radiated away. This process will continue until a stringy size is reached. However, this radiation is unable to get rid of the $\omega$ mod $k$ conserved charge carried by the initial configuration. We thus conclude that the tower of (highly unstable) winding modes decay to the twisted sector that resolves the singularity (plus radiation). Albeit indirect, this provides a connection between the tower of states relevant for the Distance Conjecture and the resolution of the ETW singularity.\footnote{When the $(D+1)$-dimensional solution is pure flat space and there is no orbifold (i.e., $k=0$), this argument simply captures the fact that all the winding modes along the circle become unstable because they can be unwrapped.} 

\medskip

Motivated by this improved understanding of the resolution of the ETW singularity above, let us generalize the analysis in \cite{Calderon-Infante:2023ler} to more than one extra dimension. For this, consider the reduction of gravity over an $n$-torus down to $D>2$ dimensions:
\begin{equation}
ds^2_{D+n}= e^{2\alpha\varphi}{ds^2_D} + e^{2\beta\varphi}\left(M_{ab}d\theta^ad\theta^b\right)\,.
\end{equation}
Here $M$ is a symmetric matrix of determinant 1. 
If we choose
\begin{equation}
    \alpha^2  = \frac{n}{(D+n-2)(D-2)} \,,\qquad \beta =-\frac{D-2}{n}\alpha\,.
\end{equation}
The $D$-dimensional effective action is
\begin{equation}
\frac{\mathcal{L}}{\sqrt{-g}} = R-(\partial\varphi)^2+\tfrac{1}{4} \text{Tr}{\partial M \partial M^{-1}}\,.    
\end{equation}
The $M$-matrix describes the coset $\SL(n,\mathbb{R})/SO(n)$ and accordingly the flows are geodesics on this space, which is known to be fully integrable. Following the strategy of \cite{Chemissany:2007fg} (see also \cite{Baines:2025upi}) we can use the fact that all geodesics are global $SO(n)$ transformations of geodesics with diagonal $M$ and the latter flows, with flat slicings $k=0$, lift to what can be called static Kasner solutions (see \cite{Apers:2022cyl} for the analogous statement in cosmological setups)
\begin{equation}
ds^2 = r^{2P_0}dr^2 - r^{2P_1}dt^2 + r^{2P_2}dx_1^2 + \ldots + r^{2P_{D+n-1}} dx_{D+n-1}^2\,.    
\end{equation}
Note that we used one notation for all spacelike coordinates different from $r$, which of course obscures what are compact vs non-compact coordinates.  This metric solves the vacuum Einstein equations if the powers $P$ obey:
\begin{align} \label{Kasner-constraints}
 P_0+1    &= \sum_{i=1}^{D+n-1} P_i\,,\\
 (P_0+1)^2&= \sum_{i=1}^{D+n-1} P_i^2\,.
\end{align}
To fit the domain wall ansatz in $D$ dimensions, we set $P_1=P_2=\cdots=P_D$. Also note that we can always set $P_0=0$ by redefining $r$, which we chose from now. 

This geometry has a curvature singularity at $r=0$ except when $P_i=0 \ \forall i$, namely flat space, or when there is only one non-vanishing $P$, which essentially recovers the solution found in \cite{Calderon-Infante:2023ler}. This can be seen from the Kretschmann scalar, which reads
\begin{equation}
    K \equiv R_{\mu \nu \rho \sigma} R^{\mu \nu \rho \sigma} = \frac{4}{r^4} \left( \sum_i P_i^2 (P_i-1)^2 + \sum_{i<j} P_i^2 P_j^2 \right) \, .
\end{equation}
Hence, for a generic geodesic in the $n$-torus moduli space, we cannot easily determine whether the flow uplifts to a healthy background or not. Let us however recall that the moduli space flows exploring each geodesic look identical from the viewpoint of the $D$-dimensional effective theory. Given that there is at least one such flow that can be UV-completed, as argued in section \ref{ss:moduli-space-flows}, either all moduli space flows in this model are UV-complete or it is impossible to formulate a sufficient singularity criterion that does not require UV information.  Even though this lies beyond the scope of this work, studying whether string theory on these static Kasner backgrounds is consistent could help elucidate whether a best-case-scenario singularity criterion is possible.\footnote{The behavior of strings on Kasner and null Kasner-like singularities has been studied in \cite{Copeland:2010yr,Madhu:2009jh,Narayan:2009pu}. Similar analyses for the case of static Kasner could help understanding the fate of these backgrounds.}

Let us also point out that the use of dualities in moduli space could help in this endeavor. To be more concrete, let us consider a Kasner solution in 10d string frame. If we perform T-duality along the direction $x^i$ then the new Kasner power $\tilde{P}_i$ is flipped in sign, $\tilde{P}_i=-P_i$, but the dilaton would run as
\begin{equation}
e^{\phi}=\text{const.}\quad \rightarrow e^{\tilde{\phi}}\sim r^{-P_i}=r^{\tilde{P}_i}\,.
\end{equation}
This means that in 10d Einstein frame, the original powers $P$ change as\footnote{Notice that the new $\tilde P_i$ do not need to satisfy \eqref{Kasner-constraints} due to the non-trivial dilaton flow.}
\begin{equation}
P_a \rightarrow P_a+\frac{1}{4}P_i\quad\text{when} \quad a\neq i\,,\quad \text{and}\quad P_i\rightarrow -\frac{3P_i}{4}\,.     
\end{equation}
This showcases how a solution for which the i'th circle shrinks in the singular region $r\to 0$ ($P_i>0$) can be traded for a solution in which it blows up ($\tilde{P}_i<0$), both in 10d Einstein and string frames. Notice that the theory runs to strong coupling and would require a further S-duality to keep control over string perturbation theory. 

Kasner solutions originate from geodesics in moduli space with only Cartan subalgebra charges. A generic geodesic lies on a U-duality orbit of those geodesics \cite{Chemissany:2007fg,Baines:2025upi}. It is natural to expect that a moduli space flow with a physical singularity should lead to another physical background after U-duality. In other words, a single moduli space flow having a physical singularity should imply the same for a whole worth of U-duality-related geodesics. Let us however point out that the T-dual solution described above might not provide the complete picture of the UV-complete uplift of the moduli space flow. The Buscher's rules \cite{Buscher:1987sk} that we used to relate different solutions only take into account the zero modes on the circle. As such, they can yield a smeared version of the uplifted solution. For instance, this has been shown to be the case for the T-dual of a Taub-NUT geometry (which yields the background of an NS5-brane smeared over a circle \cite{Gregory:1997te,Tong:2002rq}) and Witten's bubble of nothing \cite{Witten:1981gj} (which yields a smeared version of an UV-complete solution that remains mysterious \cite{Delgado:2023uqk}).

\section{ETW Singularities Outside Moduli Space}
\label{s:Flowsoutsidemoduli}
In this section we explore singular solutions of the form described in Section \ref{sec:EOM} in the presence of a non-vanishing scalar potential.\footnote{By non-vanishing potential we mean $V\neq0$ in the theory in which the solution has been reduced to a codimension-one flow. We emphasize that these can descend from solutions with $V=0$ in the higher dimensional theory that develop a non-vanishing potential only after reduction along the transverse sphere, either because of the curvature or the presence of flux-like contributions.} Our goal is to consider different well-known solutions with runing scalars and consider their truncations to a codimension-one ansatz, so that we can confront them with the different singularity criteria presented above and elucidate their validity. Our findings show (i) that Gubser's potential criterion seems to fail as a general necessary condition; (ii) that the horizon criterion can be at most sufficient but can only be applied with well-controlled asymptotics; (iii) that the Maldacena-Nu\~{n}ez criterion is satisfied for all solutions that have a known UV completion, as expected; and (iv) that the refined version of Gubser proposed around eq. \eqref{scaling-relation}, which arises as a natural (weaker) version of Gubser's criterion motivated by the universal properties of \emph{local dynamical cobordisms} \cite{Angius:2022aeq} is fulfilled by all the singular solutions with a  known UV completion.

\subsection{A Singularity Criterion from Dynamical Cobordisms} \label{ss:cobordism-criterion}

In this section, we present a novel singularity criterion based on the formalism of (local) dynamical cobordisms that we reviewed in Section \ref{ss:dynamical-cobordisms}. More precisely, this criterion arises as a \emph{geometrization} of Gubser's potential criterion by using the \emph{scaling relations} that show up in the context of dynamical cobordisms. Remarkably, this geometrization naturally leads to a slightly weaker version of Gubser's necessary condition for a \emph{good} singularity that, as we will see in Section \ref{ss:EFT-strings}, applies more generally since it correctly diagnoses a family of UV-complete singularities that are not embedded into (quasi-)AdS spacetime.

\medskip

The most natural way to formulate a singularity criterion is through a condition on the background locally around the singular locus. As emphasized in \cite{Maldacena:2000mw}, the Maldacena-Nu\~{n}ez criterion is of this type while Gubser's is not directly applicable to the singularity itself but to the dynamics generating it. However, in the context of (local) dynamical cobordisms the latter can be turned into a geometric condition on how badly the Ricci scalar diverges as a function of the field-space distance as the ETW singularity is approached. Indeed, from \eqref{dynamical-cobordism-potential}, we see that any potential bounded from above for $\phi \to \infty$ satisfies $a \geq 0$ or, equivalently,
\begin{equation} \label{eq:bound-delta}
    \delta \leq \delta_{\rm crit} \equiv 2\sqrt{\frac{d-1}{d-2}} \, .
\end{equation}
If we insist on focusing on the singular geometry---as opposed to the dynamics generating it---to motivate a necessary condition for a \emph{good} singularity, the only constraint on the curvature scalar \eqref{scaling-relation} that is compatible with Gubser's potential criterion is
\begin{equation} \label{eq:ETW-criterion}
    |R| \lesssim \exp \left( 2\sqrt{\frac{d-1}{d-2}} \, \phi \right) \quad {\rm as } \quad \phi\to\infty \, .
\end{equation}
Given this, and following the reasoning above, we would like to propose \eqref{eq:ETW-criterion} as a \emph{necessary} condition for a \emph{good} ETW singularity exploring an infinite-distance limit in field-space.

Even though we used Gubser's potential criterion to motivate \eqref{eq:ETW-criterion}, let us stress that they are not equivalent. Indeed, the bound \eqref{eq:bound-delta}, when introduced into \eqref{dynamical-cobordism-potential}, leads to
\begin{equation} \label{eq:ETW-potential-criterion}
    V \lesssim \exp \left( 2\sqrt{\frac{d-1}{d-2}} \, \phi \right) \quad {\rm as } \quad \phi\to\infty \, .
\end{equation}
That is, our criterion is actually weaker than that of Gubser when formulated in terms of the asymptotic potential. This can be traced back to the fact that all asymptotically positive potentials satisfying \eqref{eq:ETW-potential-criterion} lead to ETW singularities with $\delta = \delta_{\rm crit}$ in \eqref{scaling-relation} and for which the potential is irrelevant because  the kinetic term of $\phi$ is what dominates locally around the singularity. Hence, if we insist on having a criterion that only refers to the background and not to the dynamics, and on allowing for ETW singularities appearing in theories with $V(\phi)=0$ (as we argued for in Section \ref{sec:ModuliSpaceFlows}), we are automatically led to also include all asymptotically positive potentials satisfying \eqref{eq:ETW-potential-criterion}. We will see in Section \ref{ss:EFT-strings} that allowing for these cases is actually the right choice, as we will find UV-complete ETW singularities that violate Gubser's potential criterion but satisfy \eqref{eq:ETW-criterion} or, equivalently, \eqref{eq:ETW-potential-criterion}.

Before going on, let us stress that the potential criterion in \eqref{eq:ETW-potential-criterion} is not a bound on any potential that is allowed in quantum gravity. That is, it is not a Swampland bound on $V(\phi)$. As it is the case for Gubser's criterion, it rules out not the potential but the type ETW singularity generated by it. In fact, a potential violating \eqref{eq:ETW-potential-criterion} does arise in theories of quantum gravity. A simple example is given by the potential in massive Type IIA, which we will consider in Section \ref{ss:massive-typeIIA}.

\subsection{Klebanov-Tseytlin and Klebanov-Strassler solutions} \label{ss:KTvsKS}

The Klebanov-Tseytlin (KT) \cite{Klebanov:2000nc} and Klebanov-Strassler (KS) \cite{Klebanov:2000hb} solutions can be seen as codimension-one flows in a nearly-AdS$_5$ geometry. The string embedding is a truncation to the $\mathcal{N}=2$ subsector Type IIB compactified on $T^{1,1}$. We mainly follow the conventions of \cite{Cassani:2010na}, since they allow for a direct comparison between the KT and KS solutions. The scalar sector is given by the (real) fields $\{u,v,t,\phi, b_\Phi, b_\Omega, c_\Omega \}$. The fields $u$, $v$ and $t$ are related to the metric of the $T^{1,1}$;\footnote{The combination $4u+v$ is the overall volume and $u-v$ controls the relative size between the 4 dimensional base, $S^2\times S^2$, and the $U(1)$ fibre. Non-vanishing values of the modulus $t$ encode  non-diagonal mixing between the metric of the two 2-spheres, and is related to the deformation parameter in the deformed conifold \cite{Cassani:2010na}.}$\phi$ is the Type IIB dilaton  (with asymptotic value given by $\log(g_s)$); $b_\Phi$ and $b_\Omega$ are the axions from expanding the NSNS 2-form on a basis of left-invariant forms of the $T^{1,1}$; and $c_\Omega$ is the axion coming from the expansion of the the RR 2-form. We denote $M$ the quantized background $F_3$ flux, and $N$ the quantized $F_5$ flux. One can always express $N=k M+P$, with $k\in \mathbb{Z}$ (and $P$ and $M$ co-prime) to distinguish the part of the $F_5$ flux that corresponds to \emph{induced} 5-form flux, and can thus be eliminated by using the axion monodromy of $b_{\phi}$, from the one that cannot be canceled in that way and is associated to the number of regular D3-branes in the dual picture, $P$. \cite{Klebanov:2000hb}. 

The scalar potential in the 5d truncation takes the form \cite{Cassani:2010na}
\begin{equation}\label{eq:KTKSscalarpotential}
\begin{aligned}
V\, =\, &  +4 e^{-\frac{20}{3} u+\frac{4}{3} v}-24  e^{-\frac{14}{3} u-\frac{2}{3} v}  \cosh (t)\ +9 e^{-\frac{8}{3} u-\frac{8}{3} v} \ \sinh ^2 (t)  \\
& +9 e^{-\frac{20}{3} u-\frac{8}{3} v-\phi}\left(b_{\Omega}\right)^2+2 e^{-\frac{32}{3} u-\frac{8}{3} v}\left(3 b_{\Omega} c_{\Omega}-M b_{\Phi}+P\right)^2 \\
& +e^{-\frac{20}{3} u-\frac{8}{3} v+\phi}\left(9c_{\Omega}^2 \cosh (2 t)-6 M c_{\Omega} \sinh (2 t)+M^2 \cosh (2 t)\right).
\end{aligned}
\end{equation}
The three first terms come from the internal curvature of the $T^{1,1}$, the fourth term from the reduction of the kinetic term for $H_3$, the fifth from the RR $F_5$ (including the Chern-Simons term),\footnote{We have used here the axion monodromy of $b_\phi$ to reabsorb a the constant factor in the fifth term that corresponds to the induced $F_5$ flux.} and the sixth from the RR $F_3$-form.  Notice that all terms but the second one are non-negative definite, so solutions in which any of these contributions diverges near the singularity and dominates the scalar potential automatically violate Gubser's potential criterion.

\subsubsection*{The Klebanov-Tseytlin solution}
In the KT solution, the fields $t$, $b_\Omega$ and $c_\Omega$ are turned off,\footnote{From the point of view of the truncation above, this enhances the symmetry $SU(2)\times SU(2)\times\mathbb{Z}_2\times\mathbb{Z}_{2,R}$ of the $\mathcal{N}=2$ truncation above to an $SU(2)\times SU(2)\times\mathbb{Z}_2\times U(1)_{R}$ symmetry for this subsector.} and the metric ansatz for the codimension-one flow can be written as \cite{Krishnan:2018udc}
\begin{equation}
    ds_5^2=\dfrac{1}{z^2}\left( e^{2X(z)}dz^2+ e^{2Y(z)} \eta_{\mu\nu}dx^\mu dx^{\nu} \right) \, ,
\end{equation}
where $z$ is the radial coordinate. The AdS boundary is located at $z=0$ and the singularity at the value of $z$ for which the function\footnote{We have set an arbitrary scale $z_0=1$ that makes the argument of $\log(z)$ dimensionless for simplicity.} 
\begin{equation}
    h_{\mathrm{KT}}(z)=\dfrac{1}{8}\left( -4P + g_s M^2 - 4 g_s M^2 \log z\right)\, ,
\end{equation}
vanishes. The non-trivial field profiles take the form
\begin{equation}
    \phi=\log(g_s) , \ \ \ \   b_\Phi=\dfrac{g_s M}{3}- g_s M \log(z) ,  \ \ \ \  \  u=v=\dfrac{1}{4}\log \{ h_{\mathrm{KT}}(z)\},  \ \ \ \  X=4Y=\dfrac{2}{3}\log \{ h_{\mathrm{KT}}(z)\}\,.
\end{equation}
By inserting these solutions into the potential \eqref{eq:KTKSscalarpotential} one can see that the third and fourth terms vanish. Expanding near $h_{\mathrm{KT}}=0$ we get that the (positive) RR 5-form contribution dominates and the potential diverges as $V \simeq \mathcal{C}^2 \, h_{KT}^{-10/3}\to +\infty$, with $\mathcal{C}$ a flux dependent quantity. According to Gubser's potential criterion, this is a bad singularity, and hence it cannot be hidden by a (near-extremal) horizon that recovers the singular solution in a continuous way. Additionally, $g_{00}^{(5d)}\to 0$ but in 10d we have $g_{00}^{(10d)} \sim e^{-\frac{2}{3}(4u+v)}g_{00}^{(5d)}\to \infty$, and hence KT does not pass the Maldacena-Nu\~nez criterion. Finally, using the kinetic matrix for the scalars (see e.g. eq. (2.11) in \cite{Krishnan:2018udc}) this can be re-casted in the language of (local) dynamical cobordisms and gives $\delta=\frac{2\sqrt{30}}{3}$ and $a=-\frac{3}{2}$,\footnote{In \cite{Angius:2023xtu} these values were assigned to the Klebanov-Strassler solution, but the authors were in fact referring to the Klebanov-Tseytlin one.} hence clearly violating our ETW criterion \eqref{eq:ETW-potential-criterion}. This strongly supports interpreting the KT solution as a \emph{bad} singularity, as opposed to considering it as a \emph{good} one because it gets "resolved" by the KS solution. We will elaborate on our interpretation of "resolving" a \emph{good} singularity after reviewing the KS solution.

\subsubsection*{The Klebanov-Strassler solution}
As opposed to the KT solution, in KS all the fields in the truncation described above develop non-vanishing profiles, and one also has $P=0$. A convenient ansatz for the metric is \cite{Krishnan:2018udc}
\begin{equation}
    ds_5^2=e^{2\mathcal{X}(\tau)}d\tau^2+ e^{2\mathcal{Y}(\tau)} \eta_{\mu\nu}dx^\mu dx^{\nu}  \, ,
\end{equation}
where the AdS boundary is now located at $\tau\to +\infty$. The solutions to the scalar profiles are more conveniently written in terms of the auxiliary functions
\begin{equation}
\begin{split}
    K(\tau)=\dfrac{\left[\sinh(2\tau)-2\tau\right]^{1/3} }{2^{1/3}\, \sinh(\tau)}\,, \qquad l(\tau)=\dfrac{\left[\tau \coth(\tau)-1\right]\left[ \sinh(2\tau)-2\tau\right]}{4 \sinh^2(\tau)}\, , \\ \dfrac{d }{d\tau}\, h(\tau)=-\dfrac{16 g_s M^2}{ 81 \epsilon^{\frac{8}{3}}}\dfrac{l(\tau)}{K^2(\tau) \sinh^2(\tau)}\, ,
\end{split}
\end{equation}
where we have introduced the conifold deformation parameter $\epsilon$. $h(\tau)$ can be obtained by integrating along the $\tau$ direction numerically but has no closed expression. Nevertheless, it is enough to note that near $\tau=0$ it tends to a constant value, $a_0>0$ \cite{Klebanov:2000hb}. The scalars and the metric then take the form 
\begin{equation}
\begin{aligned}
 e^{2 u}& =\frac{3}{2} h(\tau)^{1 / 2}\,  \epsilon^{4 / 3}\,  K(\tau) \sinh (\tau)\, , \qquad  e^{2 v} =\frac{3}{2} \frac{h(\tau)^{1 / 2} \, \epsilon^{4 / 3}}{K(\tau)^2}\, , \qquad  e^{-t} =\tanh \left(\frac{\tau}{2}\right),  \\
c_{\Omega} & =\frac{M \tau}{3 \sinh (\tau)}\, , \quad
b_{\Phi}=\frac{g_s M \operatorname{coth}(\tau)}{3}(\tau \operatorname{coth}(\tau)-1)\,, \quad  b_{\Omega}=\frac{g_s M(\tau \cosh (\tau)-\sinh (\tau))}{3 \sinh { }^2(\tau)} \\
e^{2 \mathcal{X}} & =\frac{1}{4} \left(\frac{3}{2}\right)^{\frac{2}{3} }  \, h(\tau)^{\frac{4}{3}} \, \epsilon^{\frac{32}{9}} K(\tau)^{-\frac{4}{3}}\,  \sinh ^{\frac{4}{3}} (\tau)\, , \qquad e^{2 \mathcal{Y}}=\left(\frac{3}{2}\right)^{\frac{5}{3}}\, h(\tau)^{\frac{1}{3}}\,  \epsilon^{\frac{20}{9}} K(\tau)^{\frac{2}{3}} \, \sinh ^{\frac{4}{3}} (\tau)\, .
\end{aligned}
\end{equation}
By matching the solution near the AdS boundary with the KT one, one can check that $P=0$ and $\epsilon=\frac{2^{5/8}}{3^{3/4}}$. The location of the singularity is at $\tau=0$, where $e^{2 \mathcal{Y}}=0$. Inserting this solution in the scalar potential \eqref{eq:KTKSscalarpotential} and expanding near the singularity, it can be seen that the  contribution that drove the potential to $+\infty$ in the KT solution, namely the RR 5-form one, now vanishes near the singularity. The potential is now dominated by the second term in \eqref{eq:KTKSscalarpotential}, which comes from the internal curvature, and gives $V\simeq - \tilde{\mathcal{C}}^2 \, a_0^{-1} \ \tau^{-10/3}\to - \infty$, with $\tilde{\mathcal{C}}$ a flux-dependent polynomial. Thus, this fulfills Gubser's curvature criterion, as expected by the fact that the KS solution is known to be regular in 10d. The Maldacena-Nu\~nez criterion in 10d is trivially satisfied since  the solution is not singular. In terms the language of local dynamical cobordisms, this solution has $\delta=\frac{\sqrt{30}}{3}$ and $a=\frac{3}{8}$ so it also passes our criterion \eqref{eq:ETW-potential-criterion}, as expected.

\medskip

How can we reconcile the fact that KT is a \emph{bad} singularity, according to both Gubser's and our criterion \eqref{eq:ETW-potential-criterion}, with the usual statement that it is "resolved" by the KS solution? The crucial point to resolve this apparent contradiction is to be careful about the distinction between the asympotics far from the singularity versus the asymptotics near the singularity itself. It seems natural to label a singularity associated to a codimension-one flow of the type under consideration by the infinite distance point that it probes in the underlying field space. In this sense, we could define a \emph{good} singularity as one for which the field excursion followed by the singular solution can be UV completed (or IR completed in the holographic dual language) while still exploring the same infinite distance point. This is the case for UV complete singularities presented in the previous section, as well as for the KS one. As opposed to this, if a singular solution is modified near the singularity in such a way that the field trajectory is forced to explore a different infinite distance point, this cannot be considered as a "resolution" of the original singular solution, since the singularity is simply modified (even if the asymptotics far from the singularity stay the same). It is in this sense that we refer to original singular EFT solution as a \emph{bad} sinularity. This is what happens with KT, which cannot be UV completed without modifying the infinite distance point that the flow approaches. The KS flow modifies this trajectory while preserving the asymptotics far from the singularity. Thus, following the idea that the a \emph{good} singularity must be one whose UV resolution does not modify the infinite distance point explored by the EFT solution, it is not sufficient to be able to find another solution with the same asymptotics far away to the singularity and with a regular completion to identify a singularity as \emph{good}. This is the main lesson that we highlight from the KT and KS solutions.

The difference between the KT and KS solutions motivates a weaker notion of a horizon singularity criterion that also appears in the literature (see, e.g., \cite{Bena:2012ek}): the singularity can be cloaked by some black-hole horizon without requiring a near-extremal family that continuously recovers the singularity. That is, a solution with a horizon and the same asymptotics as the original singularity exists. Both for KT and KS, a family of solutions with a horizon that cloaks the singularity exist, but none of them is near-extremal, in the sense that there is no continuous limit in which the singular solution is recovered \cite{Buchel:2000ch,Buchel:2001gw,Gubser:2001ri,Mahato:2007zm,Caceres:2011zn,Aharony:2005zr,Aharony:2007vg,Buchel:2009bh,Buchel:2018bzp}. This clarifies why simply requiring the existence of a solution with a horizon does not directly diagnose whether the singularity is  \emph{good} or \emph{bad}.  

\subsection{Massive Type IIA flow} \label{ss:massive-typeIIA}

Let us turn our attention to the codimension-one flow solutions in 10d massive Type IIA constructed in \cite{Polchinski:1995df,Bergshoeff:1996ui}. This type of solution can feature two types of singular points. The first one is not truly a \emph{curvature} singularity, but rather a discontinuity.  It correspond to the location of a D8-brane, where the Romans mass $F_0>0$ discretely jumps.  We are interested in the second type of singularity that can arise due to the strong gravitational backreaction that characterizes codimension-one solutions, since it turns out to be an ETW curvature singularity. For this reason, we work with a particularly simple solution that only includes this ETW singularity. The massive Type IIA action reads
\begin{equation}
    S = \frac{1}{2\kappa^2} \int d^{10}x \sqrt{g} \bigg\{ R- \frac{1}{2} (\partial \phi)^2-\frac{1}{2}e^{\frac{5}{2} \phi}F_0^2
    \bigg\} \, ,
\end{equation}
where $\phi$ is the dilaton. We are interested in a BPS domain wall solution given by
\begin{equation}
    ds^2 = Z(x^9)^{\frac{1}{12}} \eta_{\mu \nu} dx^\mu dx^\nu \, , \quad e^{\phi} = Z(x^9)^{-\frac{5}{6}} \, , \quad Z(x^9) = -  \frac{3}{2}F_0 \, x^9 \, .
\end{equation}
Notice that the solution is well-defined only for $x^9<0$. As advanced, it hits an ETW singularity as $x^9\to 0$. We see that $\phi \to \infty$ as this singularity is approached, which in turn means that $V(\phi) \sim \exp(5\phi/2) \to + \infty$. Hence, the singularity does not satisfy Gubser's potential criterion. This ETW singularity was considered in the context of dynamical cobordisms in \cite{Buratti:2021yia,Buratti:2021fiv,Angius:2022aeq}. It satisfies the ETW scaling relation in \eqref{scaling-relation} with $\delta=5/\sqrt{2}>\delta_{\rm crit}$, which means that it does not only violate Gubser's criterion but also our (weaker) criterion \eqref{eq:ETW-criterion}.

Both Gubser's and our criterion agree that this singularity is \emph{bad}. We now argue that there is no good reason to think that this ETW singularity is UV-complete from a string theory viewpoint, providing a picture that is consistent with the diagnostics obtained from both criteria. First of all, let us remark again that $x^9=0$ is not interpreted as the position of the D8-brane, which is related to discrete jumps of $F_0$. As mentioned above, this happens in a generalization of the solution where one can decide the value of $x^9\neq 0$ where the D8-brane sits \cite{Bergshoeff:1996ui}. Therefore, the singularity at $x^9=0$ cannot be argued to be good by interpreting it as the location of a D8-brane. Even allowing for the introduction of D8-branes, the solution becomes singular at sufficiently large distances in any region of spacetime with non-vanishing Romans mass, and it is \emph{this} ETW singularity that we are interested in diagnosing. A globally well-defined solution is possible when O8-planes, fluxes or a combination thereof is included (see e.g., \cite{Apruzzi:2013yva}). In particular, the role of the O8-plane is to cap off the solution before reaching the ETW singularity. Similarly to what happened with KT vs KS solutions in Section \ref{ss:KTvsKS}, we do not interpret this as resolving this ETW singularity in the UV, but rather as a modification of the background to another one that may be UV completed. This is captured by the fact that the infinite distance point in field space probed by the original ETW singularity is never reached in any of these globally well-defined solutions. This excludes the two more natural candidates for ETW-branes making this singularity physical, namely D8-branes and O8-planes. Even though we cannot exclude the possibility of a more exotic scenario making the singularity physical, we see that the diagnosis given by these two criteria is consistent with the usual way in which globally well-defined solutions are usually built in this setup. Determining the fate of this ETW singularity requires a more in-depth string-theoretical analysis of the background. Even though $g_s\to\infty$ at the singularity, an uplift to M-theory is obstructed by the non-vanishing Romans mass. This has led to the expectation that the strongly coupled regime of massive Type IIA is never physically realized \cite{Aharony:2010af}, which would imply that the ETW singularity exploring this regime should be bad.\footnote{Alternatively, one could consider D4-branes probing the geometry as in \cite{Seiberg:1996bd}. We would like to thank Angel M. Uranga for suggesting this possibility to us.}

This massive Type IIA flow provides an interesting example in which both Gubser's and our criterion seem to correctly diagnose a \emph{bad} singularity. In the next subsection we present a family of \emph{good} ETW singularities that violate Gubser's but nevertheless are correctly diagnosed by our criterion \eqref{eq:ETW-criterion}. Before moving on, let us also point out that this ETW singularity satisfies the (strong) MN criterion. This is not in contradiction with the singularity being \emph{bad} since the MN criterion is formulated as a necessary but not necessarily sufficient condition for a good singularity.

\subsection{EFT strings and the D7-brane} \label{ss:EFT-strings}
We now turn to the study of different codimension-two solutions that implement a monodromy as they are encircled. These can be reduced along the transverse $S^1$ to yield a codimension-one object of the form we are interested in. We focus on  EFT strings \cite{Lanza:2020qmt,Lanza:2021udy, Lanza:2022zyg} and the D7 brane.

\subsubsection*{EFT string solutions}
EFT strings include a broad class of codimension-two solutions in 4d $\mathcal{N}=1$ theories that drive one or several moduli of the theory to infinite distance near their core and implement a monodromy of the corresponding axions. As we will see, they behave in the same way as the D7-brane solution in 10d, and it is generally expected that any codimension-two solution with at least minimal supersymmetry that drives the moduli towards infinite distance near its core behaves in a similar way. Since our approach in this work applies to codimension-one solutions, we first need to reduce the EFT strings along the transverse circle to a 3d codimension-one flow. This has been done in detail in \cite{Buratti:2021fiv}, so we will summarize the key steps for the one-modulus case here and refer the interested reader to the original reference for details. 

The 4d $\mathcal{N}=1$ EFT string solution with a complex scalar $S=s+ia$ (where $a$ denotes the axion and $s$ its saxionic partner) and K\"ahler potential $K=-\frac{2}{n^2}\log(S+\bar{S})$ includes a logarithmic profile for the saxion along the radial direction, $s=s_0 - \frac{q}{2 \pi }\log\left(\frac{r}{r_0}\right)$, and an axion background $a=a_0+\frac{\theta}{2\pi}q$, where $\theta\sim \theta+2\pi$ is the angular coordinate parameterizing the transverse circle. The reduction on the transverse circle to obtain the codimension-one flow yields a 3d theory that additionally includes a radion field, $\sigma$, parameterizing the size of said circle; and a non-vanishing scalar potential sourced by the non-vanishing $q$, which acts as an 'axionic' flux along the $S^1$.\footnote{Even though the original 4d $\mathcal{N}=1$ theory has vanishing scalar potential, the 3d EFT has a non-vanishing one. That is the reason why we consider EFT strings solutions, when reduced to codimension-one flows, as solutions with a non-vanishing scalar potential.} The corresponding 3d kinetic term and scalar potential take the form \cite{Buratti:2021fiv} 
\begin{equation}
    G_{ab}\partial_\mu  \phi^a \partial^\mu \phi^b= (\partial\sigma)^2+(\partial \phi)^2+e^{-2n \phi}(\partial a)^2\, , \qquad V=e^{-2\sqrt{2} \ \sigma-2n\phi}  \ \left( \dfrac{q}{2 \pi R_0 }\right)^2\, ,
\end{equation}
where we have introduced the canonically normalized $\phi=\frac{1}{n}\log(ns)$ to facilitate comparison with the D7-brane solutions later. The 3d metric can be written as 
\begin{equation}
\label{eq:metricEFTstringdynamicalcobo}
    ds_3^2= \left(\dfrac{r^2}{R_0^2} \ e^{\frac{2}{n}(\phi-\phi_0)}  \right)\left[ e^{\frac{2}{n}(\phi-\phi_0)}  \, dr^2 + \eta_{\mu \nu} dx^\mu dx^\nu \right]\, 
\end{equation}
and the non-trivial on-shell scalars take the form
\begin{equation}
\begin{aligned}
   \phi=\,&\phi_0+\dfrac{1}{n}\log \left[ 1-\dfrac{q n}{2\pi}\, e^{-n\phi_0}\, \log\left(\dfrac{r}{r_0}\right)\right]\, , \\  
   \sigma=\,&\frac{\sqrt{2}}{n^2} \log \left[1-\frac{qn}{2 \pi} \, e^{-n\phi_0}\,  \log \left(\frac{r}{r_0}\right)\right]+\sqrt{2}\log \left(\frac{r}{R_0}\right) \, .
   \end{aligned}
\end{equation}
Inserting these profiles in the potential and expanding near $r\to 0$, which is the position of the singularity, we obtain that the potential is dominated by the radion field excursion and thus goes to $V\to +\infty$. Therefore, when reduced to codimension-one flows, EFT strings violate Gubser's potential criterion. In the language of local dynamical cobordisms, these solutions have $\delta=\sqrt{8}$ and $a=0$ \cite{Angius:2022aeq}.\footnote{It can be seen from the metric \eqref{eq:metricEFTstringdynamicalcobo} that $|g_{00}|\to 0$ as $r\to 0$ so the strong MN criterion is satisfied in 3d, and similarly from the 4d EFT solution. It is however not possible to test the MN criterion generally in 10d as it depends on the UV-completion, so it does not provide particularly useful insight for this family of ETW singularities.}

\subsubsection*{The D7-brane solution}
Similarly, the D7-brane solution in 10d represents a codimension-two singularity which implementes a monodromy on the complex axio-dilaton as it is encircled (see e.g. \cite{Weigand:2018rez}). The 10d action includes the dilaton, $\Phi$, and the RR 0-form, $C_0$, which enter the Einstein-frame 10d action as (following the conventions of \cite{Angius:2022aeq})
\begin{equation}
    S_{10d}\supset\frac{M_{\mathrm{Pl,10}}^8}{2}\int d^{10}x \sqrt{-g_{10d}}\left[ -(\partial \Phi)^2 -\frac{1}{2}e^{2 \sqrt{2}\Phi}(\partial C_0)^2\right] \, . 
\end{equation}
The field profiles are $\Phi=-\frac{1}{\sqrt{2}}\log \left[ 1 - \frac{N}{2\pi}\log(r/\rho) \right]$ and $C_0=\frac{\theta N}{2\pi}$, where $N$ indicates the number of D7-branes and $\rho$ is some arbitrary length scale. The singularity is thus located at $r=0$. Reducing along the transverse circle one gets a codimension-one solution in the 9d truncation, which includes the aforementioned scalars plus a radion mode, $\omega$. The 9d, codimension-one solution for the metric takes the form 
\begin{equation}
    ds_9^2=\left(\dfrac{r^2}{R_0^2} Z(r)\right)^{1/7}\ \left( Z(r)dr+\eta_{\mu \nu} dx^\mu dx^\nu \right)\, , \quad \text{with} \quad Z(r)=1 - \frac{N}{2\pi}\log\left(\dfrac{r}{\rho}\right)\, .
\end{equation}
Similarly, the 9d kinetic term and scalar potential (sourced by the 'axionic' flux form the non-trivial $C_0$) read
\begin{equation}
      G_{ab}\partial_\mu  \phi^a \partial^\mu \phi^b= (\partial\omega)^2+(\partial\Phi)^2+\dfrac{1}{2}e^{2\sqrt{2}\Phi}(\partial C_0)^2\, , \qquad V=\dfrac{1}{2}e^{2\sqrt{2}\Phi-\frac{4\sqrt{14}}{7}\omega}\, \left(\dfrac{N}{2\pi R_0} \right)^2\, ,
\end{equation}
And the non-trivial scalar flows are
\begin{equation}
    \Phi=-\frac{1}{\sqrt{2}}\log \left[Z(r) \right]\, \qquad \omega=\dfrac{\sqrt{14}}{7} \log \left[ \dfrac{r^2}{R_0^2} Z(r)\right]\, .
\end{equation}
Similarly to the EFT strings described above, inserting the solutions for the scalars in the  9d scalar potential yields $V(r)=\frac{1}{2}\left(\frac{R_0}{r}\right)^{16/7}Z(r)^{-22/7}\left(\frac{N}{2\pi R_0}\right)^2$ and blows up to positive infinity near the singularity,  as the radion and the string coupling travel an infinite distance in field space. This means that the D7-brane, when reduced to a codimension-one flow in 9d, violates Gubser's necessary condition on the potential for a singularity to be \emph{good}.\footnote{Similarly to the case of EFT strings, it can be checked directly from the metric that the singularity at the core of the D7 fulfills the strong MN criterion both in 9d and in 10d.} It can be checked from e.g. the form of the on-shell scalar potential that the 9d solution has $\delta=4\sqrt{14}/7$ and $a=0$ \cite{Angius:2022aeq}.\footnote{The value of $a=0$ can be understood from the fact that the contribution to potential energy of the flow is dominated by the field variation $\phi^{\prime}(r)^2=g^{rr}\left(\dfrac{d\Phi}{dr}\right)^2+g^{rr}\left(\dfrac{d\omega}{dr}\right)^2$, namely $\dfrac{V(r)}{\phi^{\prime}(r)^2}\sim Z(r)^{-5/2}\to 0$ as $r\to 0$.} 

\subsubsection*{Main lessons}
We have seen that both the EFT strings and the D7-brane seem to be in tension with Gubser's potential criterion. Furthermore, it is transparent that the D7-brane is a perfectly legit solution in string theory, and there is ample evidence that EFT string solutions can be uplifted to well-behaved solutions in string theory \cite{Lanza:2021udy, Marchesano:2022avb, Marchesano:2022axe, Martucci:2022krl, Lanza:2022zyg, Marchesano:2023thx, Martucci:2024trp, Marchesano:2024tod, Hassfeld:2025uoy, Grieco:2025bjy, Monnee:2025ynn}. We take this as indication that even though Gubser's potential criterion can be formulated locally, it may not be a good diagnostic for all codimension-one solutions. In particular, for those with intricate asymptotics, as is the case of the EFT strings and the D7-brane. As hinted several times above, it seems natural to think that any purely local criterion (i.e. that can be formulated with information that can be read exclusively from the behavior of the solution near the singularity) to diagnose a potentially \emph{good} singularity should be formulated in terms of the geometry, instead of in terms of an arbitrary contribution to the on-shell action like that of the scalar potential. This idea, already mentioned in \cite{Maldacena:2000mw}, is the main motivation behind our criterion \eqref{eq:ETW-criterion}, which is remarkably fulfilled by EFT strings and the D7-brane. In fact, our proposal for a \emph{necessary} criterion for \emph{good} singularities includes all the ones that are good according to Gubser's potential criterion, but also succeeds to accommodate some that do not, such as the ones presented in this subsection.

\section{Non-extremal D$p$-branes and a Finite Temperature Distance Conjecture}
\label{s:FiniteT}
In this section we first revisit near-extremal black D$p$-brane solutions reduced to codimension-one flows along the transverse sphere in order to explore the dependence of their temperature and entropy on the field space distance as the extremal limit is approached. We find an exponential relation between the two, which motivates the formulation of a finite temperature generalization of the distance conjecture that we present at the end of the section.

\subsection{Black D$p$-branes: Solutions and Thermodynamic Scaling Relations}

A simple way to get domain wall type solutions from string theory is by reducing the black 10d D$p$-brane solutions in Type II string theory over their transverse $(8-p)$-sphere.

\subsubsection*{The 10d solutions}

The black D$p$-brane solutions with $p=0, \ldots, 6$ were first constructed in \cite{Horowitz:1991cd}. Following \cite{Kiritsis:1999ke} (see also \cite{Lu:2009yw}), we can write them in the string frame in terms of three parameters $r_0$, $L$ and $g_s$
\begin{align} \label{eq:solution}
  ds_{10}^{2} &= H(r)^{-\frac{1}{2}} \biggl( - f(r) dt^{2} + d\vec{x}^{2}_{p} \biggl) + H(r)^{\frac{1}{2}} \biggl(\frac{1}{f(r)} dr^{2} + r^{2} d\Omega_{8-p}^{2} \biggl) \, , \\
  e^{2 \phi} &= g_{s}^{2} \, H(r)^{\frac{3-p}{2}} \, , \label{eq:profile-dilaton}
\end{align}
where
\begin{equation}
  H(r) = 1 + \frac{L^{7-p}}{r^{7-p}} \, , \quad f(r) = 1 - \frac{r_0^{7-p}}{r^{7-p}} \, .
\end{equation}
We omit the field strength profile since it is not be necessary for our purposes. The parameter $r_0$ controls the deviation from extremality, and one can recover the usual extremal D$p$-brane solution in the string frame by taking $r_0\to 0$. 

This solution has a horizon placed at $r=r_0$. Relatedly, the black D$p$-brane has a temperature and entropy given by 
\begin{equation}
\begin{split}
	T &= \frac{7-p}{4 \pi} \frac{r_{0}^{(5-p)/2}}{\sqrt{r_{0}^{7-p} + L^{7-p} }} \, , \\
	S &= \frac{4 \pi \Omega_{8-p} V_p}{(2\pi)^{7}g_s^2\alpha^{\prime 4}} \, r_{0}^{\frac{9-p}{2}} \sqrt{r_{0}^{7-p} + L^{7-p} } \, ,
\end{split}
\end{equation}
where $V_p$ is the p-dimensional D$p$-brane volume and $\Omega_{8-p}$ the volume of the curvature-normalised $(8-p)$-sphere. Notice that, in string units, the temperature of extremal D5-branes is finite and that of D6-branes becomes infinite in the extremal limit. Below we will consider the temperature in different units.

The quantized D$p$-brane charge can be written in terms of $r_0$ and $L$ as
\begin{equation}
  N = \frac{(7-p) \Omega_{8-p} }{(2\pi)^{7}g_s^2\alpha^{\prime 4} \, T_p} \ L^{\frac{7-p}{2}} \sqrt{r_{0}^{7-p} + L^{7-p}} \, ,
\end{equation}
where $T_p$ is the (extremal) D$p$-brane tension. Since we are interested in working at fixed charge, it will be convenient for us to rewrite $L$ in terms of $N$ and $r_0$ as
\begin{equation}
  L^{7-p} = \sqrt{\rho^{2(7-p)} + \frac{1}{4} r_{0}^{2(7-p)}} - \frac{1}{2} r_{0}^{7-p} \, , 
\end{equation}
where we have introduced the length scale 
\begin{equation} \label{eq:rho}
  \rho^{7-p}  = \frac{(2\pi)^{7}g_s^2\alpha^{\prime 4} \, T_p \, N}{(7-p) \Omega_{8-p} } \, .
\end{equation}

\subsubsection*{Thermodynamic Scaling Relations}

We now compute how the entropy and the temperature scale with the field space distance in the extremal limit, $r_0\to 0$, for the codimension-one flows.

\medskip

First of all, let us compactify the solution on the transverse $(8-p)$-sphere down to a codimension-one running solution. Following \cite{Angius:2022aeq}, we take the compactification ansatz
\begin{equation} \label{eq:compactification-ansatz}
  \left( g_{s} e^{- \phi(r)} \right)^{1/2} ds_{10}^{2} = e^{-2 \alpha \omega(r)} ds_{p+2}^{2} + e^{2 \beta \omega(r)} R_{0}^{2} \, d\Omega_{8-p}^{2} \, ,
\end{equation}
where the $ds^2_{10}$ is the metric in 10d string frame such that the expression is in 10d Einstein frame. We take
\begin{equation}
  \alpha^{2} = \frac{8-p}{8 \,p} \, \quad \beta^{2} = \frac{p}{8(8-p)} \, ,
\end{equation}
such that the $(p+2)$-dimensional metric is in the Einstein frame and $\omega$ is canonically normalized. $R_0$ is a reference radius for the $(8-p)$-sphere. Notice that these expressions do not make sense for $p=0$. This can be tracked to the lack of Einstein frame in two dimensions, so we restrict ourselves to $p>0$ from now on. From the above equations we can relate the 10d brane solution to a domain wall in $p+2$ dimensions and find that the constant coefficient $v$ in \eqref{v-equation} corresponds to
\begin{equation}
v = \tfrac{7-p}{2}\,r_0^{7-p}\,.    
\end{equation}
So we consistently find that removing the blackening factor in 10d ($r_0=0$) corresponds to taking $v=0$ in $p+2$ dimensions.
Matching the $(8-p)$-sphere part of the metric with \eqref{eq:solution}, we obtain the profile for the radion
\begin{equation} \label{eq:profile-radion}
  \omega(r) = \sqrt{\frac{2(8-p)}{p}} \log \left( \frac{r^2}{R_0^2} H(r)^{\frac{p+1}{8}} \right) \, .
\end{equation}
Matching the rest we can also compute the $(p+2)$-dimensional metric for the running solution, which we do not present explicitly as it is not needed for our discussion.

\medskip

We are interested in the value of the scalars at the horizon, since we want to compute the field space distance in the extremal limit. Evaluating equations \eqref{eq:profile-dilaton} and \eqref{eq:profile-radion} at $r=r_0$ and taking the leading order approximation for $r_0 \ll \rho$ we find:
\begin{equation} \label{eq:fields-horizon}
\begin{split}
	\phi(r_0) &= \frac{1}{2} \log \left( g_{s}^{2} H(r_0)^{\frac{3-p}{2}} \right) \simeq \frac{1}{4} \log \left( g_{s}^{4} \left( \frac{\rho}{r_0} \right)^{(3-p)(7-p)} \right)\, , \\
	\omega(r_0) &= \sqrt{\frac{8(8-p)}{p}} \log \left( \frac{r_0}{R_0} H(r_0)^{\frac{p+1}{16}} \right) \simeq \sqrt{\frac{8-p}{32 \, p}} \log \left( \left( \frac{r_0}{R_0} \right)^{16} \left( \frac{\rho}{r_0} \right)^{(p+1)(7-p)} \right) \, .
\end{split}
\end{equation}
With this, we can compute the field space distance as $r_0 \to 0$ for $p \neq 3$ :
\begin{equation} \label{eq:field-distance}
  D_{\phi,\omega} = \int_{r_0}^{r_{\star}} dr \sqrt{\frac{1}{2} \left( \frac{d \phi}{dr} \right)^2 + \left( \frac{d \omega}{dr} \right)^2} \simeq \frac{|3-p|}{2} \sqrt{\frac{9-p}{p}} \log \left( \frac{r_\star}{r_0} \right) \, ,
\end{equation}
where we have introduced an arbitrary starting point for the field space distance, $r_\star$. Notice that the dependence on $R_0$, $\rho$ and $g_s$ drops out completely, and we recover equation (4.17) of \cite{Angius:2022aeq}. This is easy to understand, since in the extremal limit the relevant part of the metric and the dilaton profile reduces to that of the extremal D$p$-brane solution
\begin{equation}
  H(r_0) = 1 + \frac{L^{7-p}}{r_0^{7-p}} \simeq 1 + \frac{\rho^{7-p}}{r_0^{7-p}} \, .
\end{equation}
For $p=3$ the dilaton piece does not run and the distance takes the form 
\begin{equation}
D_{\omega}=\sqrt{\dfrac{10}{3}}\log\left( \dfrac{r_{\star}^2+L^2}{r_{0}^2+L^2}\right)  \, .  
\end{equation}

\medskip

The next step is to take the same limit in the expressions for the temperature and entropy. Keeping again the leading order in the $r_0 \ll \rho$ approximation  we get:
\begin{equation} \label{eq:T-S-approx}
	T \simeq \frac{7-p}{4 \pi} \, \rho^{-\frac{7-p}{2}} \, r_{0}^{\frac{5-p}{2}}  \, , \qquad 
	S \simeq \frac{4 \pi \Omega_{8-p} V_p}{(2\pi)^{7}g_s^2\alpha^{\prime 4}} \, \rho^{\frac{7-p}{2}} \, r_{0}^{\frac{9-p}{2}} \, .
\end{equation}

Before substituting equation \eqref{eq:field-distance} to get the scalings of the temperature and the entropy with the field space distance, let us be careful with the units and what is being kept constant in the extremal limit $r_0 \to 0$. First of all, note that for fixed $g_s$ we have 
\begin{equation}
  T_p \sim (\alpha^{\prime})^{-\frac{p+1}{2}} \sim M_{s}^{p+1} \, ,
\end{equation}
where we have introduced the string scale $M_s^{-1} \sim \sqrt{\alpha^{\prime}}$ without being careful about (irrelevant) numerical prefactors. Thus, keeping also $N$ fixed in equation \eqref{eq:rho}, we find 
\begin{equation}
  \rho \sim M_s^{-1} \, .
\end{equation}
Plugging this back into \eqref{eq:T-S-approx}, using equation \eqref{eq:field-distance} to recast the extremal limit $r_0 \to 0$ as the infinite field distance limit $D_\phi \to \infty$ we get
\begin{equation} \label{eq:T-S-approx2}
\begin{split}
	T &\sim M_s^{\frac{7-p}{2}} r_{0}^{\frac{5-p}{2}} \sim M_s \, \exp \left( - \frac{5-p}{|3-p|} \sqrt{\frac{p}{9-p}} \, D_{\phi,\omega}\right) \, , \\
	S &\sim V_p \, M_s^{\frac{p+9}{2}} \, r_{0}^{\frac{9-p}{2}} \sim \exp \left( - \sqrt{\frac{p(9-p)}{(3-p)^{2}}} \, D_{\phi,\omega}\right) \, ,
\end{split}
\end{equation}
Here we have also used that any dimensionless constant will be kept fixed in this limit, namely
\begin{equation}
  M_s\, r_\star \sim \mathcal O(1) \, , \quad V_p \, M_s^{\frac{p+9}{2}} \, r_{\star}^{\frac{9-p}{2}} \sim \mathcal O(1) \, .
\end{equation}
Notice that we recover exponential scalings with respect to the field distance as suggested in equation \eqref{Eequation}, yet in string units. For the entropy, which is a dimensionless quantity, this is the final answer. We find the inverse of the critical exponent $\delta$ found in \cite{Angius:2022aeq}. 
 Nevertheless, for comparison to the SDC tower, we would like to measure temperature in units of the Planck scale at the horizon. For this, we need the relation between string and lower-dimensional Planck scale when doing the compactification in \eqref{eq:compactification-ansatz}, which reads
\begin{equation}
  M_{\text{Pl}}^{p} \sim M_{s}^{8} g_s^2\exp\left(-2  \phi  \right)  R_{0}^{8-p} \exp \left( \sqrt{\frac{p(8-p)}{8}} \, \omega  \right) \, .
\end{equation}
To clarify, the first exponential comes from the dilatonic coupling in 10d string frame, while the second one is the volume of the compact dimensions in the ansatz \eqref{eq:compactification-ansatz}. Taking $R_0 \sim M^{-1}_{\text{Pl},10}$\footnote{$M_{\text{Pl},10}$ is the 10d Planck mass and $R_0$ is naturally defined in the 10d Einstein frame, so it is natural to keep $R_0 M_{\text{Pl},10}$ fixed.} and $g_s$ to be fixed, we get the scaling
\begin{equation}
  \frac{M_s}{M_{\text{Pl}}} \sim \exp \left( \frac{1}{4} \phi - \sqrt{\frac{8-p}{8 \, p}} \, \omega  \right) \, .
\end{equation}
Next, we evaluate this quantity at the horizon. Plugging the expressions in \eqref{eq:fields-horizon} to get the scaling with $r_0$ and then substituting equation \eqref{eq:field-distance} to express it in terms of the field distance we get:
\begin{equation}
  \frac{M_s}{M_{\text{Pl}}} \sim r_0^{-\frac{(6-p)(3-p)}{4p}} \sim \exp \left( \frac{(3-p)(6-p)}{2 |3-p| \sqrt{p(9-p)}} \, D_{\phi,\omega} \right) \, .
\end{equation}
Finally, plugging this into the scaling for the temperature in \eqref{eq:T-S-approx}, we find
\begin{equation}
  \frac{T}{M_{\text{Pl}}} \sim \exp \left( \frac{p (3 p-19)+18}{2 |3-p| \sqrt{p(9-p)} } \, D_{\phi,\omega} \right) \, ,
\end{equation}
confirming the intuition put into the guess \eqref{Eequation}
\begin{equation}
\begin{split}
  \frac{T}{M_{\text{Pl}}} \to 0 \quad &\text{for} \quad p=2,4,5 \, , \\
  \frac{T}{M_{\text{Pl}}} \to \infty \quad &\text{for} \quad p=1,6 \, .
\end{split}
\end{equation}

In Appendix \ref{sec:AppendixBlackDpbranes} we verify the thermodynamic scaling relations directly in 10d instead of viewing branes as co-dimension one objects in $p+2$ dimensions and find that the only brane for which the temperature blows up in units of the cutoff (species scale) seems to be the D6 brane. 

\subsection{A Finite Temperature Distance Conjecture}
\begin{figure}
    \centering
    \includegraphics[width=0.5\linewidth]{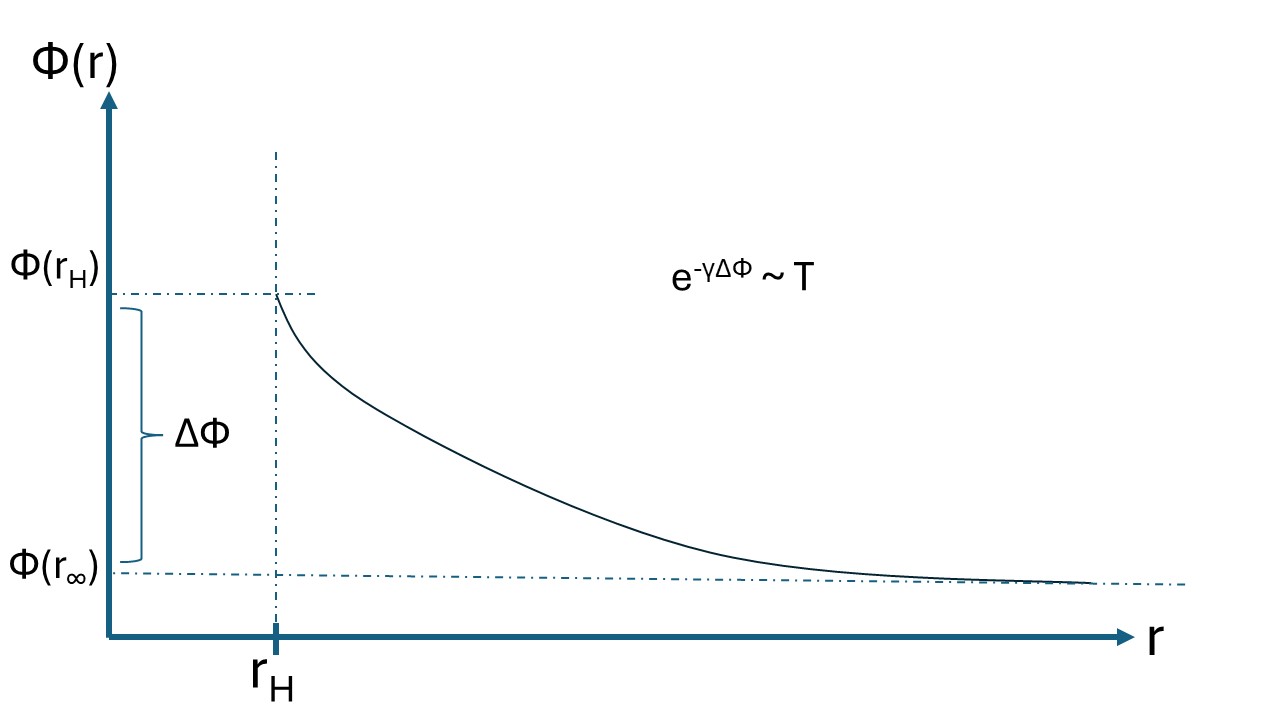}
    \caption{The flow of the scalars towards the horizon at finite temperature.}
    \label{fig:finiteT}
\end{figure}

The analysis of the black brane solutions shows that the distance in scalar field space, $\Delta_\phi$, traversed from a generic point until the horizon obeys \eqref{Eequation}:
\begin{equation}
\label{eq:FiniteTDC}
T \sim e^{-\gamma\Delta_\phi} \,,   
\end{equation}
with $\gamma$ an order one number in Planck units. This is depicted in Figure \ref{fig:finiteT}, 
which shows that if a singular flow can be cloaked by a near-extremal horizon of the kind described in \cite{Gubser:2000nd} (i.e., the ones required by Gubser's \emph{horizon} criterion), the scalar field excursion to the horizon travels a finite value and the infinite-distance limit is recovered in the extremal limit. We observe that this exponential dependence of $T$ on $\Delta_\phi$ holds for codimension-one black D$p$-branes reduced on the transverse sphere, but depending on the value of $p$ the temperature can diverge or vanish as the singular limit is approached (as is well known for D$p$-branes \cite{Kiritsis:1999ke}). Let us for now focus on branes whose temperature vanishes in the extremal limit. The physical picture that supports the exponential relation in this case could be the following;  at finite temperature the modes in the tower get excited, and this should be visible in the scalar field profile as it informs us about the excited tower modes.  At finite T the distance travelled should indeed stop since one cannot lower the energy more as there is a fixed temperature excitement. Hence, at finite $T$ we can roughly state that the $m$ mass-scale in the Distance Conjecture is rather extended to $E = m +T^c$. Thus,  it is natural to propose $E \sim e^{-\tilde{\alpha}\,\Delta_\phi}$, with $E\simeq m$ and $\tilde{\alpha}=\alpha$ (i.e. $\alpha$ the usual order-one Distance Conjecture parameter)  at zero temperature (for particles at rest), whereas $E\simeq T^{c}$ at finite temperature and $\gamma=\tilde{\alpha}/c$ giving the order-one parameter in \eqref{eq:FiniteTDC}.  According to the Emergent String Conjecture \cite{Lee:2019wij}, the light towers near infinite distance limits are either Kaluza-Klein (KK) modes or light string oscillators. We expect these to lead to distinct thermodynamics. A complete analysis of this picture lies beyond the scope of this work, but we see this as a first step towards clarifying whether the thermodynamics of ETW-branes can inform about the nature of the towers that become light in the corresponding infinite distance limits.

\section{Summary of Results and Conclusions}
\label{s:Conclusions}
We examined singular gravitational scalar flows that reach infinite distances in field space against multiple singularity criteria. These backgrounds have been widely used in recent Swampland analyses \cite{Buratti:2021yia,Buratti:2021fiv,Angius:2022aeq,Blumenhagen:2022mqw,Angius:2022mgh,Blumenhagen:2023abk,Calderon-Infante:2023ler,Angius:2023xtu,Huertas:2023syg,Angius:2023uqk,Angius:2024zjv,Huertas:2024mvy,Angius:2024pqk}. In particular, they played a key role in the bottom-up argument for the Distance Conjecture in \cite{Calderon-Infante:2023ler}. The diagnostics we applied include Gubser's  \emph{potential} and \emph{horizon} criteria \cite{Gubser:2000nd}, the Maldacena--Nu\~{n}ez criterion \cite{Maldacena:2000mw}, and a novel \emph{dynamical cobordism} criterion that we proposed in section \ref{ss:cobordism-criterion}. 
\medskip

We first worked with running solutions on moduli spaces, both with Minkowski\footnote{In these cases, even though the spacetime geometry approaches Minkowski asympotically, the scalar flow does not stop. In other words, the scalar does not approach a finite v.e.v. at asymptotic infinity.} and AdS asymptotics, which have often been set aside because their singularities were viewed as pathological. In the EFT these flows share a common form: scalars follow moduli space geodesics and infinite field-space distance is explored as an ETW singularity is approached (at finite spacetime distance). In section~\ref{ss:moduli-space-vs-criteria} we confronted this type of ETW singularities with the singularity criteria. 

First, we explicitly showed that such flows do not satisfy Gubser's \emph{horizon} criterion, which partly accounts for their prior exclusion. On the other hand, all of these ETW singularities pass Gubser's \emph{potential} criterion. To confront moduli space flows against the Maldacena--Nu\~{n}ez criterion, we considered two representative string-theoretic examples with AdS asymptotics---namely the “Gubser-Flow’’ in $AdS_5\times S^5$ \cite{Gubser:1999pk} and flows in $AdS_3\times S^3\times T^4$ \cite{Astesiano:2022qba}---and found they pass it in 10d. We also highlighted the potential frame dependence of the Maldacena--Nu\~{n}ez criterion, which arises due to the fact that a given codimension-one solution in a lower-dimensional EFT can admit various uplifts to different 10d stringy completions. Interestingly, all moduli space flows satisfy the \emph{strong} Maldacena--Nu\~{n}ez criterion when applied in the duality frame in which the internal manifold (partially) decompactifies as the infinite distance limit is explored near the ETW singularity, which is guaranteed to exist by the Emergent String Conjecture \cite{Lee:2019wij}.\footnote{When the internal space does not change along the flow, or when the latter is 10-dimensional to begin with, the uplift (or lack thereof) does not affect the behaviour of the singularity and the strong MN criterion is also satisfied.}

In section \ref{ss:moduli-space-UV-complete} we revisited the example in \cite{Calderon-Infante:2023ler} of a moduli space flow that uplifts to an orbifold singularity in one more dimension, which is a perfectly sensible background in string theory. Given that no moduli space flow satisfies Gubser's strong horizon criterion, the latter should be understood as a sufficient but not necessary condition for a \emph{good} ETW singularity. Passing it justifies keeping a flow, failing it should not imply discarding it. Motivated by this example of a UV-complete flow in moduli space, we extended the analysis to a broader family of flows in toroidal compactification that uplift to what we called static Kasner solutions. In this class of solutions, the validity of flows exploring different geodesics in moduli space relates to the question of whether the uplifted solutions are valid string theory backgrounds despite generically having a curvature singularity, which is the case when the Kasner-like uplift includes more than one shrinking circle or any circle blowing up.  We believe this last question is a concrete and addressable one, whose answer would have strong implications for our understanding of moduli space flows, but lies beyond the scope of the present work.
\medskip

We then confronted codimension-one singular flows with non-vanishing scalar potentials against the aforementioned criteria, and drew different lessons from each example. First, in section \ref{ss:KTvsKS} we studied the Klebanov-Tseytlin (KT) \cite{Klebanov:2000nc} and Klebanov-Strassler (KS) \cite{Klebanov:2000hb} solutions as codimension-one flows in 5d. The former violates Gubser's \emph{potential} criterion, whereas the latter passes it by modifying the field excursion towards the ETW singularity. This motivates interpreting KS not as a \emph{resolution} of the KT singularity, which cannot be resolved according to Gubser, but as a modification instead. More generally, and motivated by swampland insights, we propose that an ETW singularity of the kind studied here is to be thought as \emph{good} singularity that can be UV completed if the infinite distance point explored by the EFT flow is the one that is resolved in the UV. Conversely, it is not resolved if a UV-complete solution, even when it shares the same IR asymptotics, modifies or obstructs the infinite distance point explored by the original solution, as KS does. 

We also analyzed EFT-strings \cite{Lanza:2020qmt,Lanza:2021udy,Lanza:2022zyg} and D7-branes in section \ref{ss:EFT-strings}. Both of these are codimension-two solutions that can be reduced to codimension-one ETW singularities by reducing them along their transverse circle. Surprisingly, both of them fail Gubser's \emph{potential} criterion due to divergent potential towards the core of the flow. Given the ample evidence in favor of EFT strings as UV complete solutions \cite{Marchesano:2022avb, Marchesano:2022axe, Martucci:2022krl, Marchesano:2023thx, Martucci:2024trp, Marchesano:2024tod, Hassfeld:2025uoy, Grieco:2025bjy, Monnee:2025ynn}, let alone the D7-brane, we conclude that Gubser's \emph{potential} criterion, even though it can be formulated in terms of local quantities near the singularity,  is too strong as a necessary condition (at least outside AdS asymptotics). This, together with the universality of local dynamical cobordisms to describe this ETW singular flows and the idea that a \emph{good} singularity can be diagnosed from local information lead us to formulate our criterion in section \ref{ss:cobordism-criterion}. The latter is strongly motivated by the idea that any local criterion to diagnose a potentially \emph{good} singularity should be formulated in terms of the geometry, instead of the dynamics generating it (e.g., the scalar potential in Gubser's criterion) \cite{Maldacena:2000mw}.  In the context of local dynamical cobordisms, the latter can be turned into a geometric condition on how the Ricci scalar diverges with the field-space distance as the ETW singularity is approached, hence providing the desired criterion. Let us remark that our criterion \eqref{eq:ETW-criterion} automatically accepts any singularity which fulfills Gubser's criterion, but we also showed it accommodates some that do not, such as EFT strings and D7-branes.\footnote{When translated into a constraint on the scalar potential as the ETW singularity is approached, our criterion allows certain positive divergent potentials (i.e., bounded above by the exponential with $\delta\leq \delta_{\text{crit}}$), whereas Gubser's does not accept any positive divergent potential.} It would be extremely interesting to get a deeper understanding of our criterion from confronting it against other singular flows or clarifying its implications in holographic setups. Furthermore, we emphasize our analysis here is formulated for codimension-one flows with flat slicings (i.e., $k=0$ in \eqref{eq:ansatzzerotemp}), so determining how our results are connected to analyses with different slicings (see e.g. \cite{Chemissany:2011gr, Kiritsis:2025ytb}) is also crucial and left for future work.

Additionally, we examined the massive Type IIA flow \cite{Polchinski:1995df,Bergshoeff:1996ui} in section \ref{ss:massive-typeIIA}. Focusing on the ETW curvature singularity that is generated by the strong backreaction associated to the codimension-one solution in 10d (as opposed to the location of a D8-brane itself), we found that it violates both Gubser's \emph{potential} and our criterion \eqref{eq:ETW-criterion}. Given that all globally well-defined solutions constructed in the literature include either fluxes or O8-planes to cap-off the singularity (see e.g., \cite{Apruzzi:2013yva}), we interpret this in a similar manner to the KT vs. KS resolution mentioned above, namely as a singularity that is not UV-completed but avoided by the introduction of extra ingredients, consistently with the diagnostics from the aforementioned criteria. A more detailed analysis is cumbersome since the flow drives the string coupling to infinity, and the uplift to M-theory is obstructed by the non-zero Romans mass. An alternative approach may be using D4-branes to probe the geometry \cite{Seiberg:1996bd}, but this is beyond the scope of this work.
\medskip

Finally, we focused on flows whose singularities posses a near-extremal generalization and can hence be cloaked behind a regular horizon by introducing a blackening factor. In parallel with insights from minimal black holes and their thermodynamic correspondence with towers of light states near infinite-distance limits \cite{Cribiori:2023ffn,Basile:2023blg,Basile:2024dqq,Herraez:2024kux,Herraez:2025clp,Aparici:2026xxx}, we examined the relation among temperature, field-space distance, and towers for codimension-one scalar flows. Our concrete arena, detailed in section~\ref{s:FiniteT}, involved black D$p$-branes reduced on the transverse sphere, to provide codimension-one solutions \cite{Angius:2022aeq}. These flows are subject to a non-trivial potential. Across these examples we observed $T \sim e^{-\gamma\,\Delta_\phi}\,$, which resembles a finite-temperature version of the Distance Conjecture ($m \sim e^{-\alpha\,\Delta_\phi}$) \cite{Ooguri:2006in}. The sign of the exponential, though, is not universal, since depending on the object the temperature can diverge or vanish as the singular limit is approached, as is well known for black D$p$-branes depending on the value of $p$ and the sign of their specific heat. This non-universality suggests that properties of the associated towers may control the thermodynamics of the heated-up solutions, but we leave further detailed investigation of this issue for future work.
\medskip

Our analysis is a first step towards a general framework that connects scalar flows used to reach infinite distance limits in field-space with different diagnostics that separate acceptable from pathological singularities. It also provides an entry point to clarify whether towers that become light at infinite distance in the adiabatic cases enter the thermodynamics of codimension-one flows exploring the same points at finite temperature.

\section*{Acknowledgements}
	
We are very thankful to Iosif Bena, Alberto Castellano, Elias Kiritsis, Dieter L\"ust, Carmine Montella, Angel M. Uranga, Irene Valenzuela and Matteo Zatti for illuminating discussions and very useful feedback. This work was performed in part at Aspen Center for Physics, which is supported by National Science Foundation grant PHY-2210452. The work of JC is supported by the Walter Burke Institute for Theoretical Physics at Caltech and by the U.S. Department of Energy, Office of Science, Office of High Energy Physics, under Award Number DE-SC0011632. The work of GC and TVR is supported by the KU Leuven C1 grant ZKE7799C16/25/010. GC also acknowledges support from the China Scholarship Council.

\appendix

\section{Thermodynamic Scaling Relations of Black Branes in 10d}\label{sec:AppendixBlackDpbranes}

Let us  repeat the analysis above from the 10d perspective. In this case we do not compactify down to a running solution and the field space distance is directly given by
\begin{equation}
  D_{\phi} = \frac{1}{\sqrt{2}} \int_{r_0}^{r_{\star}} dr \left| \frac{d \phi}{dr} \right| \simeq \frac{|3-p|(7-p)}{4 \sqrt{2}} \log \left( \frac{r_\star}{r_0} \right) \, ,
\end{equation}
where we have used the dilaton profile in \eqref{eq:fields-horizon}. Substituting into \eqref{eq:T-S-approx}, in this case we get
\begin{equation} \label{eq:T-S-approx10d}
\begin{split}
	T &\sim M_s^{\frac{7-p}{2}} r_{0}^{\frac{5-p}{2}} \sim M_s \, \exp \left( -\frac{2 \sqrt{2} \, (5-p)}{|3-p|(7-p)} \, D_\phi\right) \, , \\
	S &\sim V_p \, M_s^{-\frac{p+9}{2}} \, r_{0}^{\frac{9-p}{2}} \sim \exp \left( -\frac{2 \sqrt{2} \, (9-p)}{|3-p|(7-p)} \, D_\phi\right) \, .
\end{split}
\end{equation}

To go to 10d Planck units at the horizon, we use the usual relation 
\begin{equation}
  \frac{M_s}{M_{\text{Pl}}} \sim \exp \left( \frac{1}{4} \phi \right) \sim \exp \left( -\frac{\text{Sgn}(p-3)}{2 \sqrt{2}} \, D_\phi \right) \, .
\end{equation}
Notice that the $\text{Sgn}(3-p)$ just reflects the fact that for $p>3$ the solution explores weak string coupling while for $p<3$ it explores strong string coupling. Plugging this back into the previous expression we then find
\begin{equation}
  \frac{T}{M_{\text{Pl}}} \sim \exp \left( \frac{(p-2) p-19}{2 \sqrt{2} \, |3-p| (7-p) } \, D_\phi \right) \, .
\end{equation}
Looking at the sign of the exponent, in this case we find that in the extremal limit
\begin{equation}
\begin{split}
  \frac{T}{M_{\text{Pl},10}} \to 0 \quad &\text{for} \quad p=1,2,4,5 \, , \\
  \frac{T}{M_{\text{Pl}, 10}} \to \infty \quad &\text{for} \quad p=6 \, .
\end{split}
\end{equation}

We are interested in measuring the temperature in terms of the species scale, since it is the maximum cutoff of the EFT. We already concluded that in terms of the 10d string scale we have that the temperature goes to zero in the extremal limit for $p=0,\ldots 4$, remains finite for 5-branes and blows up for 6-branes.  Since $g_s\to 0$ for $p=4,5,6$, $M_s$ is the right species scale and this result is what we wanted.  On the other hand, for $p=1$ the species scale is the scale of the D1-string. Taking into acccount that
\begin{equation}
   \frac{M_{D1}}{M_{\text{Pl}}} \sim \exp \left( - \frac{1}{4} \phi \right) \sim \exp \left(- \frac{1}{2 \sqrt{2}} \, D_\phi \right) \, ,
\end{equation}
we can compute
\begin{equation}
  \frac{T}{M_{D1}} \sim \frac{T}{M_{\text{Pl}}} \frac{M_{\text{Pl}}}{M_{D1}} \sim \exp \left(- \frac{1}{3 \sqrt{2}} \, D_\phi \right) \to 0 \, .
\end{equation}
In conclusion, the temperature for $p=1$ in units of the species scale goes to zero.

Finally, for $p=0,2$ the species scale corresponds to the 11d Planck scale. We have
\begin{equation}
    \frac{M_{11}}{M_{\text{Pl}}} \sim g_{s}^{-1/12} \sim \exp \left( - \frac{1}{6\sqrt{2}} D_\phi \right) \, .
\end{equation}
Thus, for $p=0$ we have
\begin{equation}
  \frac{T}{M_{11}} \sim \frac{T}{M_{\text{Pl}}} \frac{M_{\text{Pl}}}{M_{11}} \sim \exp \left(- \frac{\sqrt{2}}{7} \, D_\phi \right) \to 0 \, .
\end{equation}
Similarly, for the $p=2$ we have
\begin{equation}
  \frac{T}{M_{11}} \sim \frac{T}{M_{\text{Pl}}} \frac{M_{\text{Pl}}}{M_{11}} \sim \exp \left(- \frac{13\sqrt{2}}{15} \, D_\phi \right) \to 0 \, .
\end{equation}
In conclusion, for $p=0,2$ the temperature also goes to zero when compared to the species scale. Hence, the only brane for which the temperature blows up in units of the cutoff seems to be the D6.

\bibliography{ref.bib}

\providecommand{\href}[2]{#2}\begingroup\raggedright\begin{thebibliography}{10}

\bibitem{McNamara:2019rup}
J.~McNamara and C.~Vafa, {\it {Cobordism Classes and the Swampland}},  \href{http://arxiv.org/abs/1909.10355}{{\tt arXiv:1909.10355}}.

\bibitem{Buratti:2021yia}
G.~Buratti, M.~Delgado, and A.~M. Uranga, {\it {Dynamical tadpoles, stringy cobordism, and the SM from spontaneous compactification}},  {\em JHEP} {\bf 06} (2021) 170, [\href{http://arxiv.org/abs/2104.02091}{{\tt arXiv:2104.02091}}].

\bibitem{Buratti:2021fiv}
G.~Buratti, J.~Calder\'on-Infante, M.~Delgado, and A.~M. Uranga, {\it {Dynamical Cobordism and Swampland Distance Conjectures}},  {\em JHEP} {\bf 10} (2021) 037, [\href{http://arxiv.org/abs/2107.09098}{{\tt arXiv:2107.09098}}].

\bibitem{Angius:2022aeq}
R.~Angius, J.~Calder\'on-Infante, M.~Delgado, J.~Huertas, and A.~M. Uranga, {\it {At the end of the world: Local Dynamical Cobordism}},  {\em JHEP} {\bf 06} (2022) 142, [\href{http://arxiv.org/abs/2203.11240}{{\tt arXiv:2203.11240}}].

\bibitem{Blumenhagen:2022mqw}
R.~Blumenhagen, N.~Cribiori, C.~Kneissl, and A.~Makridou, {\it {Dynamical cobordism of a domain wall and its companion defect 7-brane}},  {\em JHEP} {\bf 08} (2022) 204, [\href{http://arxiv.org/abs/2205.09782}{{\tt arXiv:2205.09782}}].

\bibitem{Angius:2022mgh}
R.~Angius, M.~Delgado, and A.~M. Uranga, {\it {Dynamical Cobordism and the beginning of time: supercritical strings and tachyon condensation}},  {\em JHEP} {\bf 08} (2022) 285, [\href{http://arxiv.org/abs/2207.13108}{{\tt arXiv:2207.13108}}].

\bibitem{Blumenhagen:2023abk}
R.~Blumenhagen, C.~Kneissl, and C.~Wang, {\it {Dynamical Cobordism Conjecture: solutions for end-of-the-world branes}},  {\em JHEP} {\bf 05} (2023) 123, [\href{http://arxiv.org/abs/2303.03423}{{\tt arXiv:2303.03423}}].

\bibitem{Calderon-Infante:2023ler}
J.~Calder\'on-Infante, A.~Castellano, A.~Herr\'aez, and L.~E. Ib\'a\~nez, {\it {Entropy bounds and the species scale distance conjecture}},  {\em JHEP} {\bf 01} (2024) 039, [\href{http://arxiv.org/abs/2306.16450}{{\tt arXiv:2306.16450}}].

\bibitem{Angius:2023xtu}
R.~Angius, J.~Huertas, and A.~M. Uranga, {\it {Small Black Hole Explosions}},  \href{http://arxiv.org/abs/2303.15903}{{\tt arXiv:2303.15903}}.

\bibitem{Huertas:2023syg}
J.~Huertas and A.~M. Uranga, {\it {Aspects of Dynamical Cobordism in AdS/CFT}},  \href{http://arxiv.org/abs/2306.07335}{{\tt arXiv:2306.07335}}.

\bibitem{Angius:2023uqk}
R.~Angius, A.~Makridou, and A.~M. Uranga, {\it {Intersecting end of the world branes}},  {\em JHEP} {\bf 03} (2024) 110, [\href{http://arxiv.org/abs/2312.16286}{{\tt arXiv:2312.16286}}].

\bibitem{Angius:2024zjv}
R.~Angius, {\it {End of the world brane networks for infinite distance limits in CY moduli space}},  {\em JHEP} {\bf 09} (2024) 178, [\href{http://arxiv.org/abs/2404.14486}{{\tt arXiv:2404.14486}}].

\bibitem{Huertas:2024mvy}
J.~Huertas and A.~M. Uranga, {\it {End of the world brane dynamics in holographic 4d $ \mathcal{N} $ = 4 SU(N) with 3d $ \mathcal{N} $ = 2 boundary conditions}},  {\em JHEP} {\bf 01} (2025) 002, [\href{http://arxiv.org/abs/2410.05368}{{\tt arXiv:2410.05368}}].

\bibitem{Angius:2024pqk}
R.~Angius, A.~M. Uranga, and C.~Wang, {\it {End of the world boundaries for chiral quantum gravity theories}},  {\em JHEP} {\bf 03} (2025) 064, [\href{http://arxiv.org/abs/2410.07322}{{\tt arXiv:2410.07322}}].

\bibitem{Apers:2025pon}
F.~Apers, M.~Montero, and I.~Valenzuela, {\it {Backtracking AdS flux vacua}},  \href{http://arxiv.org/abs/2506.03314}{{\tt arXiv:2506.03314}}.

\bibitem{Apers:2026lgi}
F.~Apers, {\it {On the DGKT brane dual and its decoupling}},  \href{http://arxiv.org/abs/2601.15093}{{\tt arXiv:2601.15093}}.

\bibitem{Raucci:2026fzp}
S.~Raucci, I.~Ruiz, and I.~Valenzuela, {\it {Alice in Warpland: KK modes, Warped Compactifications and the Swampland}},  \href{http://arxiv.org/abs/2603.11163}{{\tt arXiv:2603.11163}}.

\bibitem{Nevoa:2025xiq}
V.~Nevoa, S.~Raman, and C.~Vafa, {\it {Elementary Constituents Conjecture}},  \href{http://arxiv.org/abs/2511.13813}{{\tt arXiv:2511.13813}}.

\bibitem{Gubser:2000nd}
S.~S. Gubser, {\it {Curvature singularities: The Good, the bad, and the naked}},  {\em Adv. Theor. Math. Phys.} {\bf 4} (2000) 679--745, [\href{http://arxiv.org/abs/hep-th/0002160}{{\tt hep-th/0002160}}].

\bibitem{Maldacena:2000mw}
J.~M. Maldacena and C.~Nunez, {\it {Supergravity description of field theories on curved manifolds and a no go theorem}},  {\em Int. J. Mod. Phys. A} {\bf 16} (2001) 822--855, [\href{http://arxiv.org/abs/hep-th/0007018}{{\tt hep-th/0007018}}].

\bibitem{Klebanov:2000nc}
I.~R. Klebanov and A.~A. Tseytlin, {\it {Gravity duals of supersymmetric SU(N) x SU(N+M) gauge theories}},  {\em Nucl. Phys. B} {\bf 578} (2000) 123--138, [\href{http://arxiv.org/abs/hep-th/0002159}{{\tt hep-th/0002159}}].

\bibitem{Klebanov:2000hb}
I.~R. Klebanov and M.~J. Strassler, {\it {Supergravity and a confining gauge theory: Duality cascades and chi SB resolution of naked singularities}},  {\em JHEP} {\bf 08} (2000) 052, [\href{http://arxiv.org/abs/hep-th/0007191}{{\tt hep-th/0007191}}].

\bibitem{Lanza:2020qmt}
S.~Lanza, F.~Marchesano, L.~Martucci, and I.~Valenzuela, {\it {Swampland Conjectures for Strings and Membranes}},  {\em JHEP} {\bf 02} (2021) 006, [\href{http://arxiv.org/abs/2006.15154}{{\tt arXiv:2006.15154}}].

\bibitem{Lanza:2021udy}
S.~Lanza, F.~Marchesano, L.~Martucci, and I.~Valenzuela, {\it {The EFT stringy viewpoint on large distances}},  {\em JHEP} {\bf 09} (2021) 197, [\href{http://arxiv.org/abs/2104.05726}{{\tt arXiv:2104.05726}}].

\bibitem{Lanza:2022zyg}
S.~Lanza, F.~Marchesano, L.~Martucci, and I.~Valenzuela, {\it {Large Field Distances from EFT strings}},  {\em PoS} {\bf CORFU2021} (2022) 169, [\href{http://arxiv.org/abs/2205.04532}{{\tt arXiv:2205.04532}}].

\bibitem{Marchesano:2022avb}
F.~Marchesano and M.~Wiesner, {\it {4d strings at strong coupling}},  {\em JHEP} {\bf 08} (2022) 004, [\href{http://arxiv.org/abs/2202.10466}{{\tt arXiv:2202.10466}}].

\bibitem{Martucci:2022krl}
L.~Martucci, N.~Risso, and T.~Weigand, {\it {Quantum gravity bounds on $ \mathcal{N} $ = 1 effective theories in four dimensions}},  {\em JHEP} {\bf 03} (2023) 197, [\href{http://arxiv.org/abs/2210.10797}{{\tt arXiv:2210.10797}}].

\bibitem{Marchesano:2022axe}
F.~Marchesano and L.~Melotti, {\it {EFT strings and emergence}},  {\em JHEP} {\bf 02} (2023) 112, [\href{http://arxiv.org/abs/2211.01409}{{\tt arXiv:2211.01409}}].

\bibitem{Marchesano:2023thx}
F.~Marchesano, L.~Melotti, and L.~Paoloni, {\it {On the moduli space curvature at infinity}},  {\em JHEP} {\bf 02} (2024) 103, [\href{http://arxiv.org/abs/2311.07979}{{\tt arXiv:2311.07979}}].

\bibitem{Martucci:2024trp}
L.~Martucci, N.~Risso, A.~Valenti, and L.~Vecchi, {\it {Wormholes in the axiverse, and the species scale}},  {\em JHEP} {\bf 07} (2024) 240, [\href{http://arxiv.org/abs/2404.14489}{{\tt arXiv:2404.14489}}].

\bibitem{Marchesano:2024tod}
F.~Marchesano, L.~Melotti, and M.~Wiesner, {\it {Asymptotic curvature divergences and non-gravitational theories}},  {\em JHEP} {\bf 02} (2025) 151, [\href{http://arxiv.org/abs/2409.02991}{{\tt arXiv:2409.02991}}].

\bibitem{Hassfeld:2025uoy}
B.~Hassfeld, J.~Monnee, T.~Weigand, and M.~Wiesner, {\it {Emergent strings in Type IIB Calabi-Yau compactifications}},  {\em JHEP} {\bf 01} (2026) 140, [\href{http://arxiv.org/abs/2504.01066}{{\tt arXiv:2504.01066}}].

\bibitem{Grieco:2025bjy}
A.~Grieco, I.~Ruiz, and I.~Valenzuela, {\it {EFT strings and dualities in 4d $\mathcal{N}=1$}},  \href{http://arxiv.org/abs/2504.16984}{{\tt arXiv:2504.16984}}.

\bibitem{Monnee:2025ynn}
J.~Monnee, T.~Weigand, and M.~Wiesner, {\it {Physics and geometry of complex structure limits in type IIB Calabi-Yau compactifications}},  {\em JHEP} {\bf 03} (2026) 063, [\href{http://arxiv.org/abs/2509.07056}{{\tt arXiv:2509.07056}}].

\bibitem{Charmousis:2010zz}
C.~Charmousis, B.~Gouteraux, B.~S. Kim, E.~Kiritsis, and R.~Meyer, {\it {Effective Holographic Theories for low-temperature condensed matter systems}},  {\em JHEP} {\bf 11} (2010) 151, [\href{http://arxiv.org/abs/1005.4690}{{\tt arXiv:1005.4690}}].

\bibitem{Ooguri:2006in}
H.~Ooguri and C.~Vafa, {\it {On the Geometry of the String Landscape and the Swampland}},  {\em Nucl. Phys. B} {\bf 766} (2007) 21--33, [\href{http://arxiv.org/abs/hep-th/0605264}{{\tt hep-th/0605264}}].

\bibitem{Cribiori:2023ffn}
N.~Cribiori, D.~L{\"u}st, and C.~Montella, {\it {Species entropy and thermodynamics}},  {\em JHEP} {\bf 10} (2023) 059, [\href{http://arxiv.org/abs/2305.10489}{{\tt arXiv:2305.10489}}].

\bibitem{Basile:2023blg}
I.~Basile, D.~L{\"u}st, and C.~Montella, {\it {Shedding black hole light on the emergent string conjecture}},  \href{http://arxiv.org/abs/2311.12113}{{\tt arXiv:2311.12113}}.

\bibitem{Basile:2024dqq}
I.~Basile, N.~Cribiori, D.~L{\"u}st, and C.~Montella, {\it {Minimal Black Holes and Species Thermodynamics}},  \href{http://arxiv.org/abs/2401.06851}{{\tt arXiv:2401.06851}}.

\bibitem{Herraez:2024kux}
A.~Herr{\'a}ez, D.~L{\"u}st, J.~Masias, and M.~Scalisi, {\it {On the Origin of Species Thermodynamics and the Black Hole - Tower Correspondence}},  {\em SciPost Phys.} {\bf 18} (2025) [\href{http://arxiv.org/abs/2406.17851}{{\tt arXiv:2406.17851}}].

\bibitem{Herraez:2025clp}
A.~Herr{\'a}ez, D.~L{\"u}st, J.~Masias, and C.~Montella, {\it {A short overview on the Black Hole-Tower Correspondence and Species Thermodynamics}},  in {\em {24th Hellenic School and Workshops on Elementary Particle Physics and Gravity}}, 6, 2025.
\newblock \href{http://arxiv.org/abs/2506.02335}{{\tt arXiv:2506.02335}}.

\bibitem{Aparici:2026xxx}
M.~Aparici, A.~Herraez, and J.~Masias, {\it {Black Hole-Tower Correspondence: Species backreaction in minimal black hole limits}},  {\em To appear}.

\bibitem{Kuperstein:2014zda}
S.~Kuperstein, B.~Truijen, and T.~Van~Riet, {\it {Non-SUSY fractional branes}},  {\em JHEP} {\bf 03} (2015) 161, [\href{http://arxiv.org/abs/1411.3358}{{\tt arXiv:1411.3358}}].

\bibitem{Halliwell:1986ja}
J.~J. Halliwell, {\it {Scalar Fields in Cosmology with an Exponential Potential}},  {\em Phys. Lett. B} {\bf 185} (1987) 341.

\bibitem{Hartong:2006rt}
J.~Hartong, A.~Ploegh, T.~Van~Riet, and D.~B. Westra, {\it {Dynamics of generalized assisted inflation}},  {\em Class. Quant. Grav.} {\bf 23} (2006) 4593--4614, [\href{http://arxiv.org/abs/gr-qc/0602077}{{\tt gr-qc/0602077}}].

\bibitem{Calderon-Infante:2022nxb}
J.~Calder{\'o}n-Infante, I.~Ruiz, and I.~Valenzuela, {\it {Asymptotic accelerated expansion in string theory and the Swampland}},  {\em JHEP} {\bf 06} (2023) 129, [\href{http://arxiv.org/abs/2209.11821}{{\tt arXiv:2209.11821}}].

\bibitem{Apers:2022cyl}
F.~Apers, J.~P. Conlon, M.~Mosny, and F.~Revello, {\it {Kination, meet Kasner: on the asymptotic cosmology of string compactifications}},  {\em JHEP} {\bf 08} (2023) 156, [\href{http://arxiv.org/abs/2212.10293}{{\tt arXiv:2212.10293}}].

\bibitem{Chemissany:2011gr}
W.~Chemissany, B.~Janssen, and T.~Van~Riet, {\it {Einstein Branes}},  {\em JHEP} {\bf 10} (2011) 002, [\href{http://arxiv.org/abs/1107.1427}{{\tt arXiv:1107.1427}}].

\bibitem{Witten:1981gj}
E.~Witten, {\it {Instability of the Kaluza-Klein Vacuum}},  {\em Nucl. Phys. B} {\bf 195} (1982) 481--492.

\bibitem{Skenderis:2006fb}
K.~Skenderis and P.~K. Townsend, {\it {Pseudo-Supersymmetry and the Domain-Wall/Cosmology Correspondence}},  {\em J. Phys. A} {\bf 40} (2007) 6733--6742, [\href{http://arxiv.org/abs/hep-th/0610253}{{\tt hep-th/0610253}}].

\bibitem{Lu:1996er}
H.~Lu, S.~Mukherji, and C.~N. Pope, {\it {From p-branes to cosmology}},  {\em Int. J. Mod. Phys. A} {\bf 14} (1999) 4121--4142, [\href{http://arxiv.org/abs/hep-th/9612224}{{\tt hep-th/9612224}}].

\bibitem{Basile:2023rvm}
I.~Basile and C.~Montella, {\it {Domain walls and distances in discrete landscapes}},  {\em JHEP} {\bf 02} (2024) 227, [\href{http://arxiv.org/abs/2309.04519}{{\tt arXiv:2309.04519}}].

\bibitem{Basile:2022zee}
I.~Basile, {\it {Emergent Strings at an Infinite Distance with Broken Supersymmetry}},  {\em Astronomy} {\bf 2} (2023), no.~3 206--225, [\href{http://arxiv.org/abs/2201.08851}{{\tt arXiv:2201.08851}}].

\bibitem{Debusschere:2024rmi}
C.~Debusschere, F.~Tonioni, and T.~Van~Riet, {\it {A distance conjecture beyond moduli?}},  \href{http://arxiv.org/abs/2407.03715}{{\tt arXiv:2407.03715}}.

\bibitem{Mohseni:2024njl}
A.~Mohseni, M.~Montero, C.~Vafa, and I.~Valenzuela, {\it {On measuring distances in the quantum gravity landscape}},  {\em JHEP} {\bf 12} (2024) 168, [\href{http://arxiv.org/abs/2407.02705}{{\tt arXiv:2407.02705}}].

\bibitem{Demulder:2024glx}
S.~Demulder, D.~Lust, and T.~Raml, {\it {Navigating string theory field space with geometric flows}},  \href{http://arxiv.org/abs/2412.10364}{{\tt arXiv:2412.10364}}.

\bibitem{Bergshoeff:2008be}
E.~Bergshoeff, W.~Chemissany, A.~Ploegh, M.~Trigiante, and T.~Van~Riet, {\it {Generating Geodesic Flows and Supergravity Solutions}},  {\em Nucl. Phys. B} {\bf 812} (2009) 343--401, [\href{http://arxiv.org/abs/0806.2310}{{\tt arXiv:0806.2310}}].

\bibitem{Lee:2019wij}
S.-J. Lee, W.~Lerche, and T.~Weigand, {\it {Emergent strings from infinite distance limits}},  {\em JHEP} {\bf 02} (2022) 190, [\href{http://arxiv.org/abs/1910.01135}{{\tt arXiv:1910.01135}}].

\bibitem{Gubser:1999pk}
S.~S. Gubser, {\it {Dilaton driven confinement}},  \href{http://arxiv.org/abs/hep-th/9902155}{{\tt hep-th/9902155}}.

\bibitem{Astesiano:2022qba}
D.~Astesiano, D.~Ruggeri, M.~Trigiante, and T.~Van~Riet, {\it {Instantons and no wormholes in $AdS_3\times S^3 \times CY_2$}},  {\em Phys. Rev. D} {\bf 105} (2022), no.~8 086022, [\href{http://arxiv.org/abs/2201.11694}{{\tt arXiv:2201.11694}}].

\bibitem{Chemissany:2007fg}
W.~Chemissany, A.~Ploegh, and T.~Van~Riet, {\it {A Note on scaling cosmologies, geodesic motion and pseudo-susy}},  {\em Class. Quant. Grav.} {\bf 24} (2007) 4679--4690, [\href{http://arxiv.org/abs/0704.1653}{{\tt arXiv:0704.1653}}].

\bibitem{Baines:2025upi}
S.~Baines, V.~Collazuol, B.~Fraiman, M.~Gra{\~n}a, and D.~Waldram, {\it {The Boundary of Symmetric Moduli Spaces and the Swampland Distance Conjecture}},  \href{http://arxiv.org/abs/2508.18401}{{\tt arXiv:2508.18401}}.

\bibitem{Copeland:2010yr}
E.~J. Copeland, G.~Niz, and N.~Turok, {\it {The string wave function across a Kasner singularity}},  {\em Phys. Rev. D} {\bf 81} (2010) 126006, [\href{http://arxiv.org/abs/1001.5291}{{\tt arXiv:1001.5291}}].

\bibitem{Madhu:2009jh}
K.~Madhu and K.~Narayan, {\it {String spectra near some null cosmological singularities}},  {\em Phys. Rev. D} {\bf 79} (2009) 126009, [\href{http://arxiv.org/abs/0904.4532}{{\tt arXiv:0904.4532}}].

\bibitem{Narayan:2009pu}
K.~Narayan, {\it {Null cosmological singularities and free strings}},  {\em Phys. Rev. D} {\bf 81} (2010) 066005, [\href{http://arxiv.org/abs/0909.4731}{{\tt arXiv:0909.4731}}].

\bibitem{Buscher:1987sk}
T.~H. Buscher, {\it {A Symmetry of the String Background Field Equations}},  {\em Phys. Lett. B} {\bf 194} (1987) 59--62.

\bibitem{Gregory:1997te}
R.~Gregory, J.~A. Harvey, and G.~W. Moore, {\it {Unwinding strings and t duality of Kaluza-Klein and h monopoles}},  {\em Adv. Theor. Math. Phys.} {\bf 1} (1997) 283--297, [\href{http://arxiv.org/abs/hep-th/9708086}{{\tt hep-th/9708086}}].

\bibitem{Tong:2002rq}
D.~Tong, {\it {NS5-branes, T duality and world sheet instantons}},  {\em JHEP} {\bf 07} (2002) 013, [\href{http://arxiv.org/abs/hep-th/0204186}{{\tt hep-th/0204186}}].

\bibitem{Delgado:2023uqk}
M.~Delgado, {\it {The bubble of nothing under T-duality}},  {\em JHEP} {\bf 05} (2024) 333, [\href{http://arxiv.org/abs/2312.09291}{{\tt arXiv:2312.09291}}].

\bibitem{Cassani:2010na}
D.~Cassani and A.~F. Faedo, {\it {A Supersymmetric consistent truncation for conifold solutions}},  {\em Nucl. Phys. B} {\bf 843} (2011) 455--484, [\href{http://arxiv.org/abs/1008.0883}{{\tt arXiv:1008.0883}}].

\bibitem{Krishnan:2018udc}
C.~Krishnan, H.~Raj, and P.~N. Bala~Subramanian, {\it {On the KKLT Goldstino}},  {\em JHEP} {\bf 06} (2018) 092, [\href{http://arxiv.org/abs/1803.04905}{{\tt arXiv:1803.04905}}].

\bibitem{Bena:2012ek}
I.~Bena, A.~Buchel, and O.~J.~C. Dias, {\it {Horizons cannot save the Landscape}},  {\em Phys. Rev. D} {\bf 87} (2013), no.~6 063012, [\href{http://arxiv.org/abs/1212.5162}{{\tt arXiv:1212.5162}}].

\bibitem{Buchel:2000ch}
A.~Buchel, {\it {Finite temperature resolution of the Klebanov-Tseytlin singularity}},  {\em Nucl. Phys. B} {\bf 600} (2001) 219--234, [\href{http://arxiv.org/abs/hep-th/0011146}{{\tt hep-th/0011146}}].

\bibitem{Buchel:2001gw}
A.~Buchel, C.~P. Herzog, I.~R. Klebanov, L.~A. Pando~Zayas, and A.~A. Tseytlin, {\it {Nonextremal gravity duals for fractional D-3 branes on the conifold}},  {\em JHEP} {\bf 04} (2001) 033, [\href{http://arxiv.org/abs/hep-th/0102105}{{\tt hep-th/0102105}}].

\bibitem{Gubser:2001ri}
S.~S. Gubser, C.~P. Herzog, I.~R. Klebanov, and A.~A. Tseytlin, {\it {Restoration of chiral symmetry: A Supergravity perspective}},  {\em JHEP} {\bf 05} (2001) 028, [\href{http://arxiv.org/abs/hep-th/0102172}{{\tt hep-th/0102172}}].

\bibitem{Mahato:2007zm}
M.~Mahato, L.~A. Pando~Zayas, and C.~A. Terrero-Escalante, {\it {Black Holes in Cascading Theories: Confinement/Deconfinement Transition and other Thermal Properties}},  {\em JHEP} {\bf 09} (2007) 083, [\href{http://arxiv.org/abs/0707.2737}{{\tt arXiv:0707.2737}}].

\bibitem{Caceres:2011zn}
E.~Caceres, C.~Nunez, and L.~A. Pando-Zayas, {\it {Heating up the Baryonic Branch with U-duality: A Unified picture of conifold black holes}},  {\em JHEP} {\bf 03} (2011) 054, [\href{http://arxiv.org/abs/1101.4123}{{\tt arXiv:1101.4123}}].

\bibitem{Aharony:2005zr}
O.~Aharony, A.~Buchel, and A.~Yarom, {\it {Holographic renormalization of cascading gauge theories}},  {\em Phys. Rev. D} {\bf 72} (2005) 066003, [\href{http://arxiv.org/abs/hep-th/0506002}{{\tt hep-th/0506002}}].

\bibitem{Aharony:2007vg}
O.~Aharony, A.~Buchel, and P.~Kerner, {\it {The Black hole in the throat: Thermodynamics of strongly coupled cascading gauge theories}},  {\em Phys. Rev. D} {\bf 76} (2007) 086005, [\href{http://arxiv.org/abs/0706.1768}{{\tt arXiv:0706.1768}}].

\bibitem{Buchel:2009bh}
A.~Buchel, {\it {Hydrodynamics of the cascading plasma}},  {\em Nucl. Phys. B} {\bf 820} (2009) 385--416, [\href{http://arxiv.org/abs/0903.3605}{{\tt arXiv:0903.3605}}].

\bibitem{Buchel:2018bzp}
A.~Buchel, {\it {Klebanov-Strassler black hole}},  {\em JHEP} {\bf 01} (2019) 207, [\href{http://arxiv.org/abs/1809.08484}{{\tt arXiv:1809.08484}}].

\bibitem{Polchinski:1995df}
J.~Polchinski and E.~Witten, {\it {Evidence for heterotic - type I string duality}},  {\em Nucl. Phys. B} {\bf 460} (1996) 525--540, [\href{http://arxiv.org/abs/hep-th/9510169}{{\tt hep-th/9510169}}].

\bibitem{Bergshoeff:1996ui}
E.~Bergshoeff, M.~de~Roo, M.~B. Green, G.~Papadopoulos, and P.~K. Townsend, {\it {Duality of type II 7 branes and 8 branes}},  {\em Nucl. Phys. B} {\bf 470} (1996) 113--135, [\href{http://arxiv.org/abs/hep-th/9601150}{{\tt hep-th/9601150}}].

\bibitem{Apruzzi:2013yva}
F.~Apruzzi, M.~Fazzi, D.~Rosa, and A.~Tomasiello, {\it {All AdS$_7$ solutions of type II supergravity}},  {\em JHEP} {\bf 04} (2014) 064, [\href{http://arxiv.org/abs/1309.2949}{{\tt arXiv:1309.2949}}].

\bibitem{Aharony:2010af}
O.~Aharony, D.~Jafferis, A.~Tomasiello, and A.~Zaffaroni, {\it {Massive type IIA string theory cannot be strongly coupled}},  {\em JHEP} {\bf 11} (2010) 047, [\href{http://arxiv.org/abs/1007.2451}{{\tt arXiv:1007.2451}}].

\bibitem{Seiberg:1996bd}
N.~Seiberg, {\it {Five-dimensional SUSY field theories, nontrivial fixed points and string dynamics}},  {\em Phys. Lett. B} {\bf 388} (1996) 753--760, [\href{http://arxiv.org/abs/hep-th/9608111}{{\tt hep-th/9608111}}].

\bibitem{Weigand:2018rez}
T.~Weigand, {\it {F-theory}},  {\em PoS} {\bf TASI2017} (2018) 016, [\href{http://arxiv.org/abs/1806.01854}{{\tt arXiv:1806.01854}}].

\bibitem{Horowitz:1991cd}
G.~T. Horowitz and A.~Strominger, {\it {Black strings and P-branes}},  {\em Nucl. Phys. B} {\bf 360} (1991) 197--209.

\bibitem{Kiritsis:1999ke}
E.~Kiritsis and T.~R. Taylor, {\it {Thermodynamics of D-brane probes}},  {\em PoS} {\bf trieste99} (1999) 027, [\href{http://arxiv.org/abs/hep-th/9906048}{{\tt hep-th/9906048}}].

\bibitem{Lu:2009yw}
J.~X. Lu and S.~Roy, {\it {Remarks on the instability of black Dp-branes}},  {\em Phys. Lett. B} {\bf 686} (2010) 254--258, [\href{http://arxiv.org/abs/0911.3341}{{\tt arXiv:0911.3341}}].

\bibitem{Kiritsis:2025ytb}
E.~Kiritsis, S.~Morales-Tejera, and C.~Rosen, {\it {de Sitter versus Anti de Sitter flows and the (super)gravity landscape: Part II}},  \href{http://arxiv.org/abs/2510.12373}{{\tt arXiv:2510.12373}}.

\end{thebibliography}\endgroup
\bibliographystyle{JHEP}
	
\end{document}